\documentclass[a4paper,11pt]{article}
\pdfoutput=1 

\usepackage{jcappub} 

\usepackage[T1]{fontenc}
\usepackage{amssymb}
\usepackage{amsmath}
\usepackage{amsfonts}
\usepackage{epsfig}
\usepackage{bm} 
\usepackage{color}
\usepackage[dvipsnames]{xcolor}
\usepackage{comment}
\usepackage{float}
\usepackage[utf8]{inputenc}
\usepackage[normalem]{ulem}
\usepackage{graphics}

\usepackage[capitalise]{cleveref}

\renewcommand{\i}{\textrm i}

\definecolor{bostonuniversityred}{rgb}{0.8, 0.0, 0.0}

\title{Probing multi-step electroweak phase transition with multi-peaked primordial gravitational waves spectra}

\author[a]{Ant\'onio~P.~Morais}
\author[b]{Roman~Pasechnik}

\affiliation[a]{Departamento de F\'\i sica, Universidade de Aveiro and CIDMA,\\ 
	Campus de Santiago, 3810-183 Aveiro, Portugal}
\affiliation[b]{Department of Astronomy and Theoretical Physics, Lund University,\\ 
	221 00 Lund, Sweden}

\emailAdd{aapmorais@ua.pt}
\emailAdd{roman.pasechnik@thep.lu.se}

\abstract{Multi-peaked spectra of the primordial gravitational waves are considered as a phenomenologically
	relevant source of information about the dynamics of sequential phase transitions in the early Universe.
	In particular, such signatures trace back to specific patterns of the first-order electroweak phase 
	transition in the early Universe occurring in multiple steps. Such phenomena appear to be rather
	generic in multi-scalar extensions of the Standard Model. In a particularly simple extension of 
	the Higgs sector, we have identified and studied the emergence of sequential long- and short-lasting 
	transitions as well as their fundamental role in generation of multi-peaked structures in the primordial 
	gravitational-wave spectrum.
	We discuss the potential detectability of these signatures by the proposed gravitational-wave 
	interferometers.}

\begin{document}

\maketitle
\flushbottom	
	

\section{Introduction}
\label{Sect:Intro}

With the long-awaited discovery of the Higgs boson at the Large Hadron Collider (LHC) \citep{Aad:2012tfa,Chatrchyan:2012xdj},
the particle content of the Standard Model (SM) has finally been completed but also the question about the accessibility of new
phenomena beyond-the-SM (BSM) becomes more and more precious. Currently, the absence of new physics indications either suggests 
that new particles and/or interactions can only show up at a larger energy scale beyond the current reach of collider measurements, 
or is due to a lack of sensitivity of the current measurements to very rare phenomena. Clearly, the greater challenge in probing such new
phenomena means a weaker interplay and interactions between the SM and new physics sectors indicating a growing demand in 
new methods and tools.

The recent major discovery of a binary neutron star merger in astrophysics via the gravitational-wave (GWs) channel has opened
a new era of multi-messenger astronomy (for a detailed review on cosmological GW sources, see e.g.~Refs.~\citep{Maggiore:2018sht,Caprini:2018mtu} and references therein). 
It is also being considered as a novel experimental window into the new physics related violent phenomena that have been possibly occurring in the very early Universe
such as the first-order phase transitions (for a recent thorough discussion, see e.g.~Refs.~\cite{Mazumdar:2018dfl,Hashino:2018wee}). There is a further big potential
in improving the corresponding sensitivities at the future space-based interferometers such as Laser Interferometer Space Antenna (LISA) experiment 
\citep{Audley:2017drz}, (Ultimate) DECi-hertz Interferometer Gravitational wave Observatory ((ultimate-)DECIGO) 
\citep{Seto:2001qf,Kudoh:2005as,Kawamura:2011zz,Kuroyanagi:2014qaa}, Big Bang Observer (BBO) BBO \citep{Crowder:2005nr,Corbin:2005ny} facilities. 
Such prospects would potentially provide an access to a plethora of 
new studies with interconnections between cosmology and particle physics (see e.g.~Refs.~\citep{Huang:2016cjm,No:2011fi,
Grojean:2006bp,Apreda:2001us,Hashino:2016rvx,Hashino:2016xoj,Kakizaki:2015wua,Dev:2016feu,Dev:2016hxv,Addazi:2018nzm,Vieu:2018zze,Angelescu:2018dkk,
Alanne:2019bsm,Addazi:2019dqt,Mohamadnejad:2019vzg,Alves:2018jsw,Alves:2018oct,Chao:2017ilw,Bian:2019szo,Dev:2019njv,Wang:2019pet}). 
Such violent processes in the early Universe as the cosmological first-order phase transitions (FOPTs) produce a stochastic GW background
via e.g.~expanding and colliding vacuum bubbles \citep{Kosowsky:1991ua,Kosowsky:1992rz} (for detailed analysis of the GW radiation induced by thermal transitions, 
see e.g.~Refs.~\citep{Hindmarsh:2013xza,Hindmarsh:2015qta} and references therein). The measurements of the corresponding GWs signals 
may be considered as a gravitational probe for BSM scenarios complementary to collider searches.

The renowned Sakharov conditions for baryogenesis \citep{Sakharov:1967dj} are among the basic motivations for considering the strongly first-order electroweak phase 
transitions (EW PTs) in the course of the thermal evolution of the Universe. Indeed, in addition to baryon number, C/CP violation, a strong departure from thermal 
equilibrium such as via FOPTs is necessary to prevent dilution of the generated baryon asymmetry. In fact, neither a sufficient CP violation nor a strong enough EW PT are generated in the framework of the SM. 
For the purpose of resolving this problem, often one introduces extended scalar sectors which typically contain additional EW Higgs doublets and singlets. Quite notably, even such minimal BSM scenarios such as the Two-Higgs Doublet Model \citep{Branco:2011iw} and the Singlet-Extended SM \citep{Barger:2007im,Barger:2008jx} enable us to successfully satisfy the Sakharov conditions giving sufficient means for the EW baryogenesis mechanism \citep{Chala:2016ykx,Vaskonen:2016yiu,Beniwal:2017eik,Cline:2012hg,Kurup:2017dzf,Li:2014wia,Jiang:2015cwa,Basler:2016obg,Basler:2017uxn,Dorsch:2013wja,Ginzburg:2010wa}. 

The additional EW doublet, singlet or even triplet scalar fields, see recent work in~\citep{Chala:2018opy} for the latter case, dramatically affect the vacuum structure which exhibits a quickly growing complexity. Due to this fact, already in simple SM extensions a possibility for PT patterns with several successive first-order transitions steps emerges. The latter sequential transitions become a rather common feature in the parameter space of such models and thus deserve a special attention. Multi-step EW PTs were previously discussed in e.g.~Refs.~\citep{Chala:2018opy,Bian:2017wfv,Chao:2017vrq,Patel:2012pi,Inoue:2015pza,Blinov:2015sna,Vaskonen:2016yiu,Ramsey-Musolf:2017tgh}, particularly, in the context of baryogenesis and also in \citep{Huang:2017laj} for the 3-3-1 model.
Normally, if phase transitions are of first order already at tree level they remain strong at higher orders as well (for more details, see 
e.g.~Refs.~\citep{Ashoorioon:2009nf,Vaskonen:2016yiu,Alanne:2016wtx,Kang:2017mkl,Vieu:2018nfq}. The FOPTs amplify the free-energy release 
thus substantially enhancing the GW signals associated with expanding vacuum bubbles of a new phase, while a possible connection between 
the observable GWs spectrum and the efficiency of baryogenesis remains questionable. Besides the strong FOPTs, there are also relatively weak and 
long-lasting EWPTs which are typically of the second order at tree level. As soon as radiative corrections are incorporated, a barrier between the two minima 
appears turning the second-order PTs into the weakly first-order ones \citep{Vieu:2018nfq,Vieu:2018zze}. The current work is devoted to a thorough analysis of 
a possible interplay between the weak and strong FOPTs in a simple scalar-sector extension with an additional Higgs $\mathrm{SU}(2)_{\rm EW}$ doublet 
and a complex scalar singlet field.

We investigate a possibility for probing the multi-scalar new physics models via their gravitational footprints that emerge due to specific sequential EWPT 
patterns and a non-trivial vacuum structure significantly extending our previous study of Ref.~\citep{Vieu:2018zze}. In order to demonstrate the basic features 
of the sequential phase transition patterns and the corresponding GW spectra, we consider a particular realisation of the Two-Higgs Doublet Model with 
an additional complex EW singlet scalar (2HDSM, in what follows). We notice that the GW spectra emerging in multi-step EWPTs in a combined pattern with, 
at least, one weak and one strong FOPTs occurring at well-separated temperatures in the early Universe naturally exhibits a multi-peaked shape. Under certain 
conditions, the latter may be, in principle, accessed by future space-based GW observatories (for an earlier discussion of multi-step transitions, 
see Ref.~\citep{Chung:2010cd}). An observation of such a characteristic signature would be a strong signal favouring multiple symmetry breaking 
stages and hence a more complicated structure of the scalar potential than the one adopted in the SM framework. We perform a sophisticated numerical 
scan over the large parts of the 2HDSM parameter space and identify regions where such multiple phase transitions could leave potentially observable GWs signatures.

The article is organised as follows. In Section~\ref{Sect:BSM} we briefly introduce the basics
of the 2HDSM structure and parameter space used in our numerical implementation. 
In Section~\ref{Sect:EWPTs}, we discuss the properties of the one-loop effective potential
and thermal corrections as well as elaborate on the formalism and examples of multi-step 
phase transitions. In Section~\ref{Sect:GWs-results}, the basic formalism and key characteristics
of the primordial GW spectrum have been described, along with numerical results of our
simulation. In Section~\ref{Sect:exotic}, we elaborate on possible exotic cosmological objects
emerging due to parallel first-order phase transition steps of different types and discuss
their consequences qualitatively. Finally, in Section~\ref{Sect:Conclusions} we provide
brief concluding remarks, while Appendix~\ref{sec:App-A} contains the basics of the GW 
production formalism in single-step transitions. 

\section{Two-Higgs doublet model with a complex singlet}
\label{Sect:BSM}

A typical new physics scenario incorporates several scalar fields in the potential that could be responsible 
for triggering the EWPTs. Even with very few scalar degrees of freedom, new very peculiar PT patterns emerge 
with multiple sequential symmetry breaking steps, see for instance, Refs.~\citep{Vieu:2018nfq,Vieu:2018zze}.
In such rather common cases, a non-trivial underlined vacuum structure and its thermal evolution is expected 
in the early Universe.

For a basic illustration of generic properties of multi-step EW FOPTs, we follow Ref.~\citep{Vieu:2018zze} and 
consider one of the minimal extensions of the SM Higgs sector that emerges as one of the possible low-energy limits
of the high-scale trinification theory studied for the first time in Refs.~\citep{Camargo-Molina:2016yqm,Camargo-Molina:2017kxd,Morais:2020odg,Morais:2020ypd,
Vieu:2018nfq}. A comprehensive analysis of the model structure and its tree-level vacuum
was performed recently in Ref.~\citep{Vieu:2018nfq}, so here we provide only a brief 
description relevant for the forthcoming discussion of EWPTs and GW signals in this model.

Besides the SM Higgs field $H_1$ and the SM gauge symmetry $G_{\rm SM}$ 
the considered 2HDSM contains an additional EW doublet $H_2$ and a complex singlet $S$ 
fields which are charged under {\rm an additional global} $\mathrm{U}(1)_{\mathrm{F}}$ 
family symmetry. The corresponding charges under $\mathrm{SU}(2)_{\rm EW}\times 
\mathrm{U}(1)_{\mathrm{Y}}\times \mathrm{U}(1)_{\mathrm{F}}$ can be summarized as follows
\begin{eqnarray}
H_1=(\bm{2},1,\tfrac{1}{6}) \,, \qquad H_2=(\bm{2},1,\tfrac{5}{6}) \,, \qquad S=(\bm{1},0,\tfrac{2}{3}) \,.
\label{eq:charges}
\end{eqnarray}
The resulting potential possesses an approximate discrete $\mathbb{Z}_2$ 
symmetry with respect to the following transformations $H_1 \to - H_1^\ast$, $H_2 \to - H_2^\ast$ 
and $S \to - S^\ast$ which significantly simplify the vacuum structure of the model. Such transformation properties together with those in Eq.~\eqref{eq:charges}, for instance the $\mathrm{U}(1)_\mathrm{F}$ charges, follow directly from the Supersymmetric Higgs Unified Trinification framework, or SHUT model for short, where the usual trinification GUT is extended with a $\mathrm{SU}(3)_\mathrm{F}$ family symmetry (for more details, see Ref.~\cite{Camargo-Molina:2017kxd}). More importantly, the referred $\mathbb{Z}_2$ symmetry is explicitly broken by soft SUSY-breaking interactions and terms of the type $a_{12s} H_1 H_2^\dagger S + \mathrm{h.c.}$ can only be generated at one-loop level induced by $\mathbb{Z}_2$-violating and $\mathrm{SU}(3)_\mathrm{F}$ conserving trilinear couplings. This means that, while $a_{12s}$ is technically allowed by the remnant $\mathrm{U}(1)_\mathrm{F}$ family symmetry, it can naturally be small in the SHUT model. This offers a well motivated simplifying argument to preserve the referred approximate $\mathbb{Z}_2$ symmetry in the 2HDSM model version that we study in this article.

At tree level, a minimal renormalizable potential with spontaneously $G_{\rm SM}\times \mathrm{U}(1)_{\mathrm{F}} \rightarrow 
\mathrm{SU}(3)_{\rm c}\times \mathrm{U}(1)_{\mathrm{e.m.}}$ breaking 
in the considered model reads
\begin{eqnarray}
V_0(H_1,H_2,S) &=& - \mu_1^2 |H_1|^{2} - \mu_2^2 |H_2|^{2} 
- \mu_s^2 |S|^{2} + \lambda_1 |H_1|^{4} + \lambda_2 |H_2|^{4} \nonumber \\
&+& \lambda_3 |H_1|^2 |H_2|^2 + \lambda_s |S|^{4} + \lambda_{s1} |H_1|^2 |S|^2 
+ \lambda_{s2} |H_2|^2 |S|^2 + \lambda'_3 (H_1H^\dagger_2) (H^\dagger_1H_2) \nonumber \\
&+& \Big(\frac12 \mu_b^2 S^2 + {\rm h.c.}\Big) \, .
\label{potential}
\end{eqnarray}
Here, the last term represents a soft breaking of $\mathrm{U}(1)_{\mathrm{F}}$ enabling to give a (small) 
pseudo-Goldstone mass to the imaginary part of $S$ field known as the Majoron and thus
making it play a role of a Dark Matter candidate \cite{Berezinsky:1993fm,Lattanzi:2007ux,Kuo:2018fgw,Lattanzi:2013uza,Bazzocchi:2008fh}. 
It is worth mentioning here that Majoron also provides an important bridge between the neutrino mass generation mechanisms and 
the characteristics of the EWPTs (and hence the resulting GWs spectra) which has been established
for the first time in Ref.~\citep{Addazi:2019dqt}. In order to generate a small pseudo-Goldstone Majoron 
mass responsible for the global $\mathrm{U}(1)_{\mathrm{F}}$ symmetry breaking, 
a possibility mentioned in Ref.~\cite{Vieu:2018nfq} is to take into account nonperturbative interactions 
of the Majoron with the gluon condensate (i.e.~via QCD anomaly), providing a conservative estimate 
$m_{S_I}^2 < 1\,{\rm MeV}^2$. Such a small Majoron mass, implying also $|\mu_b|\ll |\mu_{1,2,s}|$, 
will be safely neglected compared to masses of other particles in our numerical analysis below.

Expanding the scalar fields in terms of their real and imaginary parts
\begin{align}
\begin{aligned}
H_j &= \frac{1}{\sqrt{2}} \begin{pmatrix} \chi_j + i \chi_j' \\ 
\phi_j + h_j + i \eta_j \end{pmatrix}\,,
\end{aligned} \,\,
\begin{aligned}
S = \dfrac{1}{\sqrt{2}} ( \phi_s + S_R + i S_I)\,,
\end{aligned}
\end{align}
one determines $h_1$, $h_2$ and $S_R$ as quantum fluctuations about the classical field 
classical configurations $\phi_\alpha=\{\phi_1,\phi_2,\phi_s\}$, $\alpha=1,2,s$, respectively. 
Assume, for simplicity, that only the real component of $S$ gets a VEV such that the 
classical-field configurations in this case read
\begin{eqnarray}
H_1^T = \frac{1}{\sqrt{2}} (0,\phi_1)\,, \qquad H_2^T = \frac{1}{\sqrt{2}} (0,\phi_2)\,, \qquad 
S = \frac{\phi_s}{\sqrt{2}} \,,
\end{eqnarray}
and the classical field-dependent (tree-level) potential reads
\begin{eqnarray}
V_0(\phi_\alpha) = \sum_{\alpha} 
\Big[\frac{\lambda_{\alpha} \phi_\alpha^4}{4} - \frac{\mu_\alpha^2\phi_\alpha^2}{2} \Big] + 
\frac{\mu_b^2\phi_s^2}{2} + 
\frac{\lambda_{12} \phi_1^2\phi_2^2}{4} + 
\frac{\lambda_{s1} \phi_1^2\phi_s^2}{4} + 
\frac{\lambda_{s2} \phi_2^2\phi_s^2}{4} \,,
\label{eq:V0}
\end{eqnarray}
where $\lambda_{12} = \lambda_3 + \lambda'_3$. Defining $\overline{\lambda}_{12} = 
\lambda_3 + \theta\left(-\lambda_3^\prime\right)\lambda_3^\prime$ with $\theta(x-x_0)$ the step function centred in $x_0$, 
the classical potential is bounded from below (BFB) as long as
\begin{equation}
\begin{aligned}
&x_{12} = \overline{\lambda}_{12} + 2\sqrt{\lambda_1\lambda_2} > 0 \,,
\quad
x_{s1} = \lambda_{s1} + 2\sqrt{\lambda_1\lambda_s} > 0 \,,
\quad
x_{s2} = \lambda_{s2} + 2\sqrt{\lambda_2\lambda_s} > 0\,,
\\
&
\quad
\sqrt{\lambda_1 \lambda_2 \lambda_s} + \overline{\lambda}_{12} \sqrt{\lambda_s} + \lambda_{s1} \sqrt{\lambda_2} + \lambda_{s2} \sqrt{\lambda_1} + \sqrt{x_{12} x_{s1} x_{s2} } > 0\,,
\quad
\lambda_{1,2,s} > 0\,,  \quad
\end{aligned}
\label{BFB}
\end{equation}
are satisfied \cite{Kannike:2012pe}. We restrict all quartic couplings to be below ten in our numerical analysis, 
in consistency with a generic perturbativity constraint $|\lambda_i| < 4\pi$.

In this study, we consider only the case of softly broken $\mathrm{U}(1)_{\mathrm{F}}$ 
and spontaneously broken EW symmetries asymptotically at zero temperature, i.e.
\begin{eqnarray}
\langle \phi_s(T=0) \rangle_{\rm vac} = 0\,, \qquad 
\langle \phi_1(T=0) \rangle_{\rm vac}\equiv v_h \simeq 246.22 \,, \qquad
\langle \phi_2(T=0) \rangle_{\rm vac} = 0\,,
\end{eqnarray}
In this case, the SM vacuum stability condition $\mu_1^2 = \lambda_1 v_h^2$ implies the presence 
of a SM Higgs boson which does not mix with other CP-even scalars as the mass form is readily 
diagonal. The corresponding scalar mass spectrum reads
\begin{equation}
\begin{aligned}
& m_h^2 = 2\lambda_1 v_h^2 \simeq 125\, {\rm GeV} \,, \\
&  m_{1,2}^2 \equiv M_{s_{1}}^2 = \frac{\lambda_{12} v_h^2}{2} - \mu_2^2\,,
\quad m_{3,4}^2 \equiv M_{s_{2}}^2 = \frac{\lambda_3 v_h^2}{2} - \mu_2^2\,, \\
&  m_{S_R}^2 \equiv M_{s_{3}}^2 = \frac{\lambda_{s1} v_h^2}{2} + \mu_b^2 - \mu_s^2\,,
\quad m_{S_I}^2 = \frac{\lambda_{s1} v_h^2}{2} - \mu_b^2 - \mu_s^2  \,.
\label{mass}
\end{aligned}
\end{equation}
Here, the $\mathrm{U}(1)_{\mathrm{F}}$ breaking term, $\mu_b^2$, introduces a small splitting 
between the Majoron CP-odd scalar mass $m_{S_I}$ and one of the CP-even scalar masses 
$m_{S_R}$. Note, in the presence of $\mu_b^2>0$ the Majoron mass $m_{S_I}$ would be 
the lowest in the scalar sector. The positively-definite scalar masses squared, together with 
the BFB conditions (\ref{BFB}), ensure the vacuum stability.

In addition to the SM fermions, the model may also contain additional species of 
vector-like fermions at a TeV scale that could be relevant e.g.~for an enhanced CP violation in the model.
These additional heavy fermionic components typically play a secondary role in the properties
of the EWPTs in this model and hence their impact on the GWs spectrum is expected to be minor. 
So, for the purposes of the current pioneering study of multi-peak characteristics of the resulting 
GWs spectrum we omit such degrees of freedom in this model and are focused primarily on 
its rich scalar sector.

As was discussed in Refs.~\citep{Vieu:2018nfq,Vieu:2018zze} the main features of EWPTs in this model, 
such as sequential FOPTs, are rather generic phenomena relevant for various multi-Higgs SM extensions. 
So, the model under consideration, due to its apparent simplicity, could be viewed as an important benchmark model 
for future thorough studies of cosmological implications of multi-scalar new physics scenarios.

\section{Multi-step phase transitions in 2HDSM}
\label{Sect:EWPTs}

As the Universe expands and cools down, thermal evolution of its EW-breaking vacuum state is governed by 
the temperature-dependent part of the effective potential (see e.g.~Ref.~\citep{Quiros:1999jp}).
The shape of the effective potential is affected by thermal corrections which are determined by a given field 
content and symmetries of an underlying theory at any temperature $T$ e.g.~in the one-loop approximation.

\subsection{Effective $T$-dependent potential}
\label{Sect:Eff-pot}

For the purpose of exploring the features of EWPTs in the 2HDSM model under consideration, we 
construct the effective $T$-dependent potential to the one-loop order in perturbation theory 
in the following form \citep{Quiros:1999jp,Curtin:2016urg},
\begin{equation}
V_{\rm eff}(T) = V_0 + V^{(1)}_{\rm CW} + \Delta V(T) + V_{\rm ct}\,,
\label{eff-pot}
\end{equation}
where the tree-level (classical) part $V_0$ is given by Eq.~(\ref{eq:V0}), $V^{(1)}_{\rm CW}$ 
is the zero-temperature Coleman-Weinberg (CW) potential determined at one-loop level, $V_{\rm ct}$
is the counterterm potential, and the $\Delta V(T)$ term contains the lowest-order thermal corrections.

The CW potential in Landau gauge has the following standard form,
\begin{equation}
V_{\rm CW} = \sum_i (-1)^{F_i} n_i \frac{m_i^4(\phi_\alpha)}{64 \pi^2} \left( \log\left[ \frac{m_i^2(\phi_\alpha)}{\Lambda^2}\right] - c_i \right) \,,
\end{equation}
where $F=0(1)$ for bosons (fermions), $m_i^2(\phi_\alpha)$ is the $\phi_\alpha$-field 
dependent mass of the particle $i$, $n_i$ is the number of degrees of freedom  (d.o.f.'s)
for a given particle $i$, $\Lambda$ is a renormalization group (RG) scale and, 
in the $\overline{\rm MS}$-renormalization 
scheme, the constant $c_i$ is equal to $3/2$ for each d.o.f.~of scalars, fermions and longitudinally polarised gauge bosons, and to $1/2$ for transversely polarised gauge boson d.o.f.'s.
In fact, only heavy SM fermions and scalars have sufficiently large field-dependent masses 
to substantially contribute to the evolution of the shape of the potential in the course of thermal 
evolution of the Universe.

The choice of the RG scale $\Lambda$ in the fixed-order effective potential becomes particularly relevant when a given mass is very different from the EW VEV $v_h$. In order to reduce 
the dependence on the RG scale choice, in this case one typically employs the so-called 
RG-improved effective potential where the couplings and masses are replaced by their running 
values evaluated at the RG scale $\Lambda$. In our current analysis of EWPTs, we consider 
the scalar boson masses and nucleation temperatures that are typically not very far from
the EW scale, $v_h\simeq 246$ GeV, such that all the relevant potential parameters can 
be considered as (approximately) fixed at the RG scale and equal to the EW scale, 
i.e.~$\Lambda=v_h$ in what follows.

The thermal correction term $\Delta V(T)$ at one loop is given by \citep{Quiros:1999jp}:
\begin{equation}
\Delta V(T) = \frac{T^4}{2 \pi^2} \left\{ \sum_{b} n_b J_B\left[\frac{m_i^2(\phi_\alpha)}{T^2}\right] - \sum_{f} n_f J_F\left[\frac{m_i^2(\phi_\alpha)}{T^2}\right] \right\}\,,
\label{finite_T_correction}
\end{equation} 
where $J_B$ and $J_F$ are the thermal integrals for bosons and fermions, respectively, given by
\begin{align} \label{eq:JBJF}
J_{B/F}(y^2) = \int_0^\infty d x \, x^2 \log\left( 1 \mp \exp [ - \sqrt{x^2 + y^2}] \right)\,.
\end{align}

In the first non-trivial order of thermal expansion $\sim(m/T)^2$, the thermal corrections 
can be represented as follows
\begin{align}
\label{eq:y2}
\Delta V^{(1)}(T)|_{\rm L.O.} = \frac{T^2}{24} \left\{ {\rm Tr}\left[ M_{\alpha\beta}^2(\phi_\alpha) \right] + 
\sum_{i=W,Z,\gamma} n_i m_i^2(\phi_\alpha) + 
\sum_{i=t,b,\tau} \frac{n_i}{2} m_i^2(\phi_\alpha) \right\} \,,
\end{align}
where all the field-independent terms are dropped out. Here, $M_{\alpha\beta}$ is the 
field-dependent scalar Hessian matrix, and $n_i$ are the numbers of d.o.f's for a given 
particle $i$. In particular, for the SM vector bosons ($W, Z$ and transversely polarised 
photon $A_T\equiv \gamma$), ($\bar t,\bar b$) $t,b$ (anti)quarks and $\tau$-lepton we have
\begin{equation}
n_W = 6, \qquad n_Z = 3, \qquad n_\gamma = 2, \qquad n_{t,b} = 12, \qquad n_{\tau} = 4 \,.
\end{equation}
while for longitudinally polarised photon ($A_L$) and the scalar sector
\begin{equation}
n_s = 10, \qquad n_{A_L} = 1 \,.
\end{equation}
Appearance of $T^2$-terms in $\Delta V^{(1)}(T)$ signals a symmetry restoration 
at high temperatures. At the same time, the emergence of higher-order 
terms with possibly alternating signs in the effective potential are responsible for building an important barrier between 
the high- and low-$T$ phases. Such a barrier affects, in particular, the character of 
the corresponding phase transition capable of turning a second-order transition to a first-order one.

Since the trace of the Hessian in Eq.~(\ref{eq:y2}) is basis invariant, in practical calculations in
the leading-order ${\cal O}((m/T)^2)$ it is particularly convenient to use the gauge basis considering only diagonal elements of the scalar mass form. Therefore, the leading 
thermal corrections of order $T^2$ would affect only quadratic (in mean-fields) terms 
of the tree-level potential $V_0$ given by Eq.~\eqref{eq:V0}. In this way, they preserve the shape 
of $V_0$ and affect only the masses of the scalar fields.

The symmetry restoration due to $T^2$-terms in the effective potential usually signals 
the breakdown of perturbation theory in a close vicinity of the critical temperature. This means that an all-order resummation of higher order
contributions known as daisy (or ring) diagrams is required \cite{Dolan:1973qd,Parwani:1991gq,Arnold:1992rz,Espinosa:1995se}. 
The latter resummation is in practice achieved by adding the finite temperature corrections to the field-dependent masses entering the effective 
potential ({\ref{eff-pot}) as follows 
\begin{eqnarray}
 \mu_\alpha^2(T) = \mu_\alpha^2 + c_\alpha T^2 \,,
 \label{mu-T}
\end{eqnarray}
where $c_{\alpha}$ are found by analysing the infrared limit of the corresponding two-point 
correlation functions in the 2HDSM version under consideration:
\begin{align} 
& c_1 = \frac18 g^2 + \frac{1}{16}(g^2 + {g'}^2) + 
\frac12 \lambda_1 + \frac{1}{12} (\lambda_{12} + \lambda_3 + \lambda_{s1} ) + 
\frac14 (y_t^2 + y_b^2) + \frac{1}{12} y_{\tau}^2\,, \\
& c_2 = \frac18 g^2 + \frac{1}{16}(g^2 + {g'}^2) + 
\frac12 \lambda_2 + \frac{1}{12} (\lambda_{12} + \lambda_3 + \lambda_{s2} )\,, \quad
c_s = \frac13\lambda_s + \frac16( \lambda_{s1} + \lambda_{s2} ) \,.
\label{coeff_thermalmasses}
\end{align}
The thermal corrections are then universally introduced to the physical (field-dependent) scalar 
boson masses replacing $\{\mu_\alpha\}$ by the thermal mass terms $\{\mu_\alpha(T)\}$ given by 
Eq.~(\ref{mu-T}). Note, in calculations beyond the leading order performed below, such a simple 
form (\ref{eq:y2}) with a trace of the Hessian does not apply any longer. In this case, a full mass 
form diagonalisation procedure of the one-loop effective potential incorporating the thermal 
mass terms (\ref{mu-T}) should be implemented.

In a full analogy to the scalar sector, the temperature dependence of the vector boson masses 
at the leading-order is introduced by adding the $T^2$-corrections to the diagonal terms of 
the gauge boson mass matrix. It is worth noticing here that only longitudinally polarised states 
$\{W^+_L,W^-_L,Z_L,A_L\}$ receive thermal corrections such that their masses are obtained by
means of diagonalisation of the corrected mass form
\begin{eqnarray}
M_{\rm gauge}^{2}(\phi_{1,2};T) = M_{\rm gauge}^{2}(\phi_{1,2}) + \frac{11}{6}T^2 
\left( 
\begin{array}{cccc} 
g^2 & 0 & 0 & 0 \ \\  
0 & g^2 & 0 & 0 \ \\
0 & 0 & g^2 & 0 \ \\
0 & 0 & 0 & {g'}^2 
\end{array} 
\right) \,.
\end{eqnarray}
Here, the zero-temperature mass matrix is $M_{\rm gauge}^{2}(\phi_{1,2})$, with eigenvalues
\begin{eqnarray}
m_W^2(\phi_{1,2}) = \frac{\phi_1^2+\phi_2^2}{4} g^2 \,, \qquad
m_Z^2(\phi_{1,2}) = \frac{\phi_1^2+\phi_2^2}{4} (g^2+{g'}^2) \,.
\label{mZW}
\end{eqnarray}
While the mass of the transversely polarised photon, $m_\gamma$, 
is zero, in thermal medium the photon acquires a longitudinal polarisation $A_L$ which has 
a non-zero thermal mass. The gauge boson mass eigenvalues are given by
\begin{eqnarray}
&& m_{W_L}^2(\phi_{1,2};T) = m_W^2(\phi_{1,2}) + \frac{11}{6}g^2T^2\,, \\
&& m_{Z_L,A_L}^2(\phi_{1,2};T) = \frac{1}{2}m_Z^2(\phi_{1,2}) + 
\frac{11}{12}(g^2+{g'}^2)T^2 \pm {\cal D} \,,
\end{eqnarray}
with the field-dependent $W,Z$ boson masses given in Eq.~(\ref{mZW}), and
\begin{eqnarray}
{\cal D}^2 = \Big(\frac{1}{2}m_Z^2(\phi_{1,2}) + \frac{11}{12}(g^2+{g'}^2)T^2 \Big)^2 - 
\frac{11}{12} g^2{g'}^2 T^2 \Big( \phi_1^2 + \phi_2^2 + \frac{11}{3}T^2 \Big) \,.
\end{eqnarray}

Due to the presence of one-loop corrections at $T=0$ entering via the CW potential $V_{\rm CW}$, 
the VEVs and physical masses are shifted from their tree-level values. On the other hand,
one should ensure that the measured physical value of Higgs boson mass, $m_h\simeq 125$ GeV, 
and the Higgs VEV, $v_h\simeq 246$ GeV, are reproduced in the $T=0$ limit. For this purpose,
one introduces the counterterm potential $V_{\rm ct}$ in Eq.~(\ref{eff-pot}). Assuming for simplicity that 
the one-loop corrections to the quartic self-interaction couplings are small $\delta\lambda\ll \lambda$
for not very large variations in energy scale of the phase transitions, one can compute 
the counterterms only for the mass terms \citep{Jiang:2015cwa}. Provided that at $T=0$ only $H_1$
acquires a VEV,
\begin{align}
V_{\rm ct} = \frac{\delta\mu_1^2\phi_1^2}{2}\,, \qquad 
\delta \mu_1^2 = - \frac{1}{v_h} \left. \frac{\partial 
V^{(1)}_{\rm CW}}{\partial \phi_1} \right|_{\rm vac} \,,
\label{CT_corrections}
\end{align}
such that the tree-level mass formulas remain intact at zero temperature.

\subsection{Multi-step phase transitions}
\label{Sect:PTs}

The phase transitions are considered as dynamical processes describing certain 
non-perturbative solutions of the equations of motion. While in the high-$T$ regime these
processes are dominated by thermal jumps, at low $T$ they occur mainly through quantum tunnelling and 
are known as instantons (see e.g.~Refs.~\citep{Linde1983,Dine:1992wr}).
Both these cases are normally described by means of the same formalism which is based upon
a consideration of classical motion in Euclidean space. The corresponding classical action reads
\citep{Coleman:1977py}
\begin{equation}
\hat{S}_3(\hat{\phi},T) = 4 \pi \int_0^\infty \mathrm{d}r \, r^2 \left\{ \frac{1}{2} 
\left( \frac{\mathrm{d}\hat{\phi}}{\mathrm{d}r} \right)^2 + V_{\rm eff}(\hat{\phi},T) \right\} \,,
\end{equation}
where the full one-loop $T$-dependent effective potential $V_{\rm eff}$ is specified in Eq.~(\ref{eff-pot}) 
and is computed for a particular multi-scalar extension of the SM such as the 2HDSM scenario presented above.
Here, $\hat{\phi}$ is a particular solution of the equation of motion that is found by computing the path 
minimizing the energy of the corresponding field \citep{Coleman:1977py,Wainwright:2011kj}. 

The nucleation processes of vacuum bubbles happen effectively at $T_n$ known as the nucleation temperature.
It is found by a requirement that the probability for a single bubble nucleation per horizon volume is equal to unity,
such that
\begin{eqnarray}
\int_0^{t_n} \Gamma\, V_H(t)\,dt = \int_{T_n}^\infty \frac{dT}{T}
\Big( \frac{2\zeta M_{\rm Pl}}{T} \Big)^4 e^{-\hat{S}_3/T}={\cal O}(1) \,, 
\label{Tn-cond}
\end{eqnarray}
where $M_{\rm Pl}$ is the Planck scale, $V_H(t)$ is the volume of the cosmological horizon, and
$\zeta\sim 3\cdot 10^{-3}$, and
\begin{eqnarray}
\Gamma \sim A(T) e^{-\hat{S}_3/T}\,, \qquad A(T)={\cal O}(T^4) \,.
\end{eqnarray}
is the tunneling rate per unit time per unit volume \cite{Dine:1992wr}. The requirement (\ref{Tn-cond}) 
numerically translates to the following equation \citep{Dine:1992wr,Quiros:1999jp}
\begin{equation}
\frac{\hat{S}_3(T_n)}{T_n} \sim 140 \,.
\label{def_nucleation_temperature}
\end{equation}
It may also happen that equation \eqref{def_nucleation_temperature} does not have any solution such that 
transitions do not occur during the thermal history of the Universe 
\citep{Kurup:2017dzf}. While such a transition may still eventually occur at asymptotically large 
times and at $T=0$ via quantum tunneling, we do not discuss such cases in this work. Instead, 
we are focused only on transitions that happen at sufficiently large $T$ as long as 
nontrivial solutions of Eq.~\eqref{def_nucleation_temperature} can be found. 

One of the quantities we would like to study is the order parameter. In the case of one-Higgs-doublet SM
it reads \citep{Kuzmin:1985mm} 
\begin{eqnarray}
\frac{v_c}{T_c} \gtrsim 1 \,, \qquad v_c\equiv v_h(T_c) \,,
\label{FOPT-SM}
\end{eqnarray}
in terms of the critical temperature $T_c$, at which both minima become degenerate. 
In the context of electroweak baryogenesis this parameter quantifies the strength of FOPTs. 
For extended Higgs sectors e.g.~in the case 
of higher Higgs representations, however, it was demonstrated in Ref.~\citep{Ahriche:2014jna} that this criterion is relaxed compared to the doublet 
case, namely,
\begin{eqnarray}
\frac{v_c}{T_c} \gtrsim \eta \,, \qquad \eta<1 \,,
\label{FOPT-extSM}
\end{eqnarray}
with $\eta$ being dependent on a particular representation of the extended scalar sector. 

It was shown in Ref.~\citep{Patel:2011th} that in a generic case of the effective potential the sphaleron suppression 
criterion (\ref{FOPT-SM}) is manifestly not gauge invariant (see also Ref.~\citep{Nielsen:1975fs}). 
As was mentioned above, near the critical temperature, the quantum ${\cal O}(\hbar)$ corrections to the potential become 
as large as the tree-level contributions such that the power $\hbar$-expansion breaks down. A proper gauge-invariant 
resummation of daisy (or ring) diagrams for the effective potential in its minimum and the corresponding generalisation 
of the sphaleron suppression criterion has been performed in Ref.~\citep{Patel:2011th}.

In order to derive the properties of the EWPTs, one should analyse the tunneling probabilities 
and nucleation temperatures which require a detailed analysis of the effective potential away from its minima. 
The conventional formalism based upon the full one-loop ($T$-dependent) effective potential generically 
suffers from gauge dependence, see e.g.~Refs.~\citep{Chiang:2017zbz}. The gauge dependence has a less pronounced 
impact on the results if there is a barrier between the minima at tree level, hence, for the strong FOPTs \citep{Wainwright:2011qy,
Wainwright:2012zn,Blinov:2015vma}. Since the fully gauge-invariant formalism is not yet available, we follow the effective potential 
approach commonly adapted in the current literature and study all the possible phases and transitions between them 
in the framework of 2HDSM. An analogous study in a gauge-invariant approach outside the minima of the effective potential 
goes beyond the scope of the present analysis and is advised for future work.

In what follows, we define the order parameter in the 2HDSM under consideration as (c.f.~Ref.~\citep{Kurup:2017dzf})
\begin{equation}
\frac{\Delta v_n}{T_n} \gtrsim \eta \,, \quad \Delta v_n = |v(T_n+\delta T) - v(T_n-\delta T)| \,, 
\quad v(T)\equiv \sqrt{\sum_{\alpha=1,2} v_\alpha(T)^2} + v_s(T) \,,
\label{strongPT_nucleation}
\end{equation}
where $v_{1,2}(T)$ are the Higgs doublet $H_{1,2}$ VEV and $v_s(T)$ is the EW singlet VEV values at a given $T$, 
such that $\Delta v_n$ is the absolute value of difference between $v(T)$ computed
before and after a phase transition, with $\delta T$ taken to be sufficiently small, i.e. $\delta T \ll T_n$. 
Eq.~(\ref{strongPT_nucleation}) is somewhat different from a more standard sphaleron suppression criterion given in Eq.~(\ref{FOPT-extSM}). 
Indeed, first, we can have phases with non-zero EW-singlet (Majoron) VEV which certainly contributes to the sphaleron suppression. 
Second, the actual phase transition does not start at $T_c$, but rather at a somewhat lower $T_n$ when the bubble nucleation rate exceeds 
the rate of cosmological expansion. So, we consider the condition (\ref{strongPT_nucleation}) reflecting these two points 
as more generic and appropriate for our purposes. 

The emergence of FOPTs is practically relevant for production of GW signals potentially accessible by future GWs interferometers 
\citep{Kakizaki:2015wua,Hashino:2016rvx,Hashino:2016xoj}. In this work, we do not explicitly compute $\eta$ in Eq.~(\ref{strongPT_nucleation}), but only 
$\Delta v_n$ and $T_n$ separately. In practice, the condition (\ref{strongPT_nucleation}) is not really used for quantifying the strength of FOPTs in our analysis and 
plays a secondary, rather indicative role. Indeed, as we will notice below, in some cases transitions with smaller $\Delta v_n/T_n$ are capable of producing
further pronounced GW peaks than those with larger $\Delta v_n/T_n$. We take another, more phenomenological approach, namely, for each FOPT found in a vast 
numerical scan (see below) we calculate the peak-amplitude of the corresponding GW spectrum and compare it to the sensitivity curves, known for each of the planned and proposed next-generation GW interferometers. If the value of such peak-amplitude comes anywhere close to the sensitivity domain, we consider 
such FOPT as a ``would-be'' strong or, in fact, strong enough to yield a potentially observable GW signal. This will be quantified by the $\alpha$ parameter, introduced in the next section, that is related to the potential energy difference between the two vacua involved in a transition. As we will notice below, such ``physical'' 
cases can emerge from FOPTs with as low $\Delta v_n/T_n$ as $0.01-0.1$.

The nucleation temperature $T_n$ for a given transition can be found by using e.g.~the \texttt{CosmoTransitions} 
package~\citep{Wainwright:2011kj} which enables one to evaluate the Euclidean action $\hat{S}_3$ and thus
to analyse the PTs between the corresponding vacua. Here we are particularly interested in studying 
the sequential EW FOPTs referred to as multi-step PTs in what follows. As a result, one can expect more 
than a single transition for a given point in the model parameter space and, hence, successive nucleation 
of bubbles corresponding to physically different vacua states.

Considering now the VEVs of the scalar fields $v_\alpha\equiv\langle \phi_\alpha\rangle_{\rm vac}=\{v_1,v_2,v_s\}$,
one may identify several distinct configurations that represent the only existing phases \citep{Vieu:2018nfq}:
$(0,0,0)$, $(v_1,0,0)$, $(0,v_2,0)$, $(v_1,v_2,0)$ and $(0,0,v_s)$. In what follows, we label that as 
$[0]$, ${\cal H}_1$, ${\cal H}_2$, ${\cal H}_{12}$ and $\Phi$, respectively. At tree level, the possible FOPTs 
were found to be as follows: ${\cal H}_1 \leftrightarrow {\cal H}_2$, ${\cal H}_1 \leftrightarrow \Phi$, 
${\cal H}_2 \leftrightarrow \Phi$, ${\cal H}_{12} \leftrightarrow \Phi$. The latter occur already 
in the leading $(m/T)^2$ order in the thermal expansion. Thus, they are considered to be very strong 
also at one loop level. As was noted above, for simplicity, let us choose ${\cal H}_1$ to be a stable phase 
asymptotically at $T=0$, such that $v_1 \equiv v_h \simeq 246.22$ GeV. At finite temperatures, 
the Universe typically passes through intermediate phases corresponding to a set of non-trivial vacua 
with all VEVs $\{v_1,v_2,v_s\}$ being generically non-zero. Note, a discussion of the phase transitions between the ${\cal H}_1$, 
${\cal H}_2$ and $\Phi$ vacua is simple and illuminating in the study of multi-step EWPTs but also 
represents the basic features of a more involved scenario with a more generic EW-breaking ground 
state ${\cal H}_{12}$ at $T=0$. 

In the rest of this section, for simplicity, we consider the following two possible sequences of PTs 
to the stable vacuum state ${\cal H}_1$ asymptotically at $T=0$:
\begin{eqnarray}
({\rm I}): &\qquad& [0] \to \Phi \to {\cal H}_1 \,, \label{Eq:pat1} \\ 
({\rm II}): &\qquad& [0] \to \Phi \to {\cal H}_2 \to {\cal H}_1\,, \label{Eq:pat2}
\end{eqnarray}
where the intermediate phases can only be stable at finite temperatures and then get destabilised 
in the course of Universe expansion along certain directions in multi-dimensional field space.
Whenever the condition (\ref{def_nucleation_temperature}) corresponding to a transition $i \to j$ 
is satisfied, a bubble of phase $j$ is nucleated inside the phase $i$, at a given nucleation
temperature $T_n(i \to j)$.

The $[0] \to \Phi$ transition is unique among the other steps due to the fact that it becomes 
first-order by means of the thermal-loop corrections, while the other transitions considered so far are of strong 
first order already at tree-level. This is in the spirit of other models studied in 
e.g.~Refs.~\citep{Patel:2012pi,Inoue:2015pza,Blinov:2015sna,Vaskonen:2016yiu,Alanne:2016wtx}, 
when a weak cross-over transition at tree-level becomes a first-order transition at one-loop 
caused by cubic contributions in the $m/T$ expansion. Despite this,
$[0] \to \Phi$ is still considerably weaker than the other transitions.

Depending on the particular choice of the model parameters, one or another pattern can be 
realised. If we start from pattern (I), then the Universe cannot pass through the ${\cal H}_2$ 
phase, i.e.~pattern (II) does not occur. However, if for a given choice of the parameters 
the second pattern (II) is realised to start with, when the Universe cools down below 
$T_n({\cal H}_2 \to {\cal H}_1)$ it is in principle possible that both strong first-order transitions 
$\Phi \to {\cal H}_1$ and $\Phi \to {\cal H}_2$ can occur in parallel as long as the difference between 
the corresponding nucleation temperatures is small, i.e.~$T_n(\Phi \to {\cal H}_1) - 
T_n(\Phi \to {\cal H}_2) \lesssim \Delta T$, where $\Delta T \sim $ 10 GeV is the typical 
time scale of the bubble percolation process. Indeed, as the scalar potential evolves 
with temperature the initial phase $\Phi$ becomes unstable also along the ${\cal H}_1$ direction 
(due to disappearance of the potential barrier between the phases $\Phi$ and ${\cal H}_1$). Multi-step 
transitions can also occur if, e.g.~a potential barrier is generated in the $\mathcal{H}_1$ direction, 
producing a false $\mathcal{H}_1^\text{false}$ and a true $\mathcal{H}_1^\text{true}$ vacuum, 
such that the transition $\mathcal{H}_1^\text{false} \to \mathcal{H}_1^\text{true}$ leaves a visible gravitational footprint. 
This is possible when large corrections in the thermal $m/T$ expansion are triggered by large scalar quartic couplings.
For a recent thorough discussion of simultaneous phase transitions in a generic set-up and 
the corresponding GW signals, see Ref.~\citep{Croon:2018new}. 

The parallel transitions may occur, for example, when symmetries in the tree-level potential 
enforce the nucleation temperatures to be identical as in e.g.~Ref.~\citep{Ivanov:2017zjq}. 
In particular, different transition sequences e.g.~$\Phi \to {\cal H}_1$ and $\Phi \to {\cal H}_2$ might have occurred 
at the same cosmological time scale such that the 
``coexisting'' bubbles of different broken phases nucleate simultaneously (see below). 
In addition, even more exotic cosmological objects may emerge. For example, looking at 
the second and third steps in $[0] \to \Phi \to {\cal H}_2 \to {\cal H}_1$, which happen at 
$T_n(\Phi \to {\cal H}_2) \gtrsim T_n({\cal H}_2 \to {\cal H}_1)$, we notice that between 
$T_n(\Phi \to {\cal H}_2)$ and $T_n({\cal H}_2 \to {\cal H}_1)$, the ${\cal H}_2$-bubbles nucleate in 
the $\Phi$-phase. In the course of their expansion, at the temperature $T_n({\cal H}_2 \to {\cal H}_1)$
the ${\cal H}_1$-bubbles are being born and start to nucleate inside the ${\cal H}_2$-bubbles.
This means that the $\Phi$-phase gets populated with the ${\cal H}_2$-bubbles having also the ${\cal H}_1$-bubbles 
inside. This configuration gives rise to ``nested'' bubbles. These are typical examples of exotic objects 
that can emerge in multi-Higgs models. The single-step formalism cannot be applied in this case, 
and a more sophisticated analysis involving e.g.~${\cal H}_1{\cal H}_2$ bubble collisions should be developed.

However, in the considered extension of the Higgs sector, the 2HDSM scenario, a small 
hierarchy between the nucleation temperatures requires a significant fine-tuning between 
the model parameters as there is no symmetry that would make such an hierarchy natural. 
For large regions of the parameter space that we have explored in our numerical simulations 
with the full one-loop effective potential (see below), the typical differences between the nucleation 
temperatures for any of the two subsequent transitions are above 20 GeV. This means that, for instance, 
once the pattern (II) has been chosen to start with, no $\Phi \to {\cal H}_1$ transition happens in practice. 
Indeed, by the time the Universe cools down below $T_n({\cal H}_2 \to {\cal H}_1)$, the ${\cal H}_2$ 
bubbles are already completely percolated and no $\Phi$ phase remains. Since in our scenario 
an occurrence of simultaneous strong first-order transitions is highly unlikely, in what follows 
we are focused on distinct transition patterns that do not overlap in the course 
of cosmological expansion.

\section{Primordial gravitational waves}
\label{Sect:GWs-results}

Such violent processes in the early Universe as phase transitions are expected to leave 
a stochastic background of primordial GWs as a signature. In the first approximation, the primordial
stochastic GW background is statistically isotropic, stationary and Gaussian. Its power spectrum is given by 
the energy-density of the GW radiation per logarithmic frequency
\begin{equation}
h^2 \Omega_{\rm GW}(f) \equiv \frac{h^2}{\rho_c} \frac{\partial \rho_{\rm GW}}{\partial \log f}\,, 
\end{equation}
where $\rho_c$ is the critical energy density today. The production of GWs in the early 
Universe is usually considered to be driven by three different sources \cite{Caprini:2015zlo},
\begin{equation}
h^2 \Omega_{\rm GW} \simeq h^2 \Omega_{\rm coll} + h^2 \Omega_{\rm SW} + h^2 \Omega_{\rm MHD} \,,
\label{GW-Omega}
\end{equation}
due to collisions between the bubble walls \cite{Kosowsky:1991ua}, $\Omega_{\rm coll}$, 
the sound wave (SW) echoes generated after the phase transitions \cite{Hindmarsh:2013xza}, 
$\Omega_{\rm SW}$, and the associated magnetohydrodynamic (MHD) turbulences in the plasma 
\cite{Caprini:2009yp}, $\Omega_{\rm turb}$, respectively. Following the discussion in 
Ref.~\cite{Hindmarsh:2017gnf,Ellis:2019oqb}, we notice that the bubble wall collisions typically 
do not contribute to the GWs production processes in the class of multi-scalar
extensions of the SM under consideration. Only in a hypothetical case of runaway bubbles corresponding to the situation 
when the bubble wall undergoes unbounded acceleration, i.e. $v_b\to 1$, as $\alpha$ increases, the bubble 
wall collisions may become relevant. However, we do not consider this limit in our analysis and hence we no longer discuss the runaway bubbles and the bubble-wall 
collisions effect. 

In a recent study \cite{Caprini:2019egz} the most recent understanding of GW production from cosmological phase transitions is discussed, updating the formalism in \cite{Caprini:2015zlo}. Note that the state of the art expressions derived in \cite{Caprini:2019egz} do not account for MHD-turbulence effects due to large theoretical uncertainties. Therefore, we will only consider SW contributions in the remainder of this study.

The key quantities needed for the computation of the GWs power spectrum are the inverse time-scale 
$\beta$ of the phase transition (in units of the Hubble parameter $H$), 
\begin{equation}
\frac{\beta}{H} = T_n  \left. \frac{\partial}{\partial T} \left( \frac{\hat{S}_3}{T}\right) \right|_{T_n}\,,
\label{betaH}
\end{equation}
and the strength of the phase transition, $\alpha$, typically defined through the trace anomaly as \cite{Hindmarsh:2015qta,Hindmarsh:2017gnf}
\begin{equation}
\alpha = \frac{1}{\rho_\gamma} \Big[ V_i - V_f - \dfrac{T}{4} \Big( \frac{\partial V_i}{\partial T} - 
\frac{\partial V_f}{\partial T} \Big) \Big] \,,
\label{alpha}
\end{equation}
where $T_n$ is the nucleation temperature, $\hat{S}_3$ is the Euclidean action introduced above, 
$V_i$ and $V_f$ the values of the potential in the initial (metastable) and final (stable) phases 
of the effective potential, and
\begin{equation}
\rho_\gamma = g_* \frac{\pi^2}{30} T_n^4\,,  \qquad g_* \simeq 106.75 \,,
\end{equation}
is the energy density of the radiation medium at the bubble nucleation epoch found in terms of the number 
of relativistic d.o.f.'s.~$g_*$. For a more detailed discussion, see e.g. Refs.~\cite{Grojean:2006bp,Leitao:2015fmj,
Caprini:2015zlo,Caprini:2019egz}. Both quantities $\beta/H$ and $\alpha$ require a comprehensive knowledge of the effective 
potential $V_{\rm eff}(\phi_{\alpha};T)$ and are numerically computed using the \texttt{CosmoTransitions} 
package \cite{Wainwright:2011kj} in the 2HDSM extension of the SM under consideration. The corresponding GW 
signals for each of the contributions in \cref{GW-Omega}, are found schematically as
\begin{equation}
{\rm signal}~\sim~{\rm amplitude}~\times~{\rm spectral}~{\rm shape}~(f/f_{\rm peak}) \,,
\end{equation}
where $f$ is the GW frequency, and $f_\mathrm{peak}$ is the peak-frequency containing 
the redshift associated to the expansion of the Universe. In particular, the peak frequency expression that we use reads
\begin{equation}
	f_\mathrm{peak} = 26 \times 10^{-6} \left( \dfrac{1}{H R} \right) \left( \dfrac{T_\mathrm{n}}{100} \right) \left( \dfrac{g_\ast}{100~\mathrm{GeV}} \right)^{\tfrac{1}{6}} \mathrm{Hz}
	\label{eq:fpeak}
\end{equation}
where
\begin{equation}
	H R = \dfrac{H}{\beta} \left( 8 \pi \right)^{\tfrac{1}{3}} \max\left(v_b, c_s\right)
	\label{eq:HR}
\end{equation}
with $R$ the mean bubble separation and $c_s = 1/\sqrt{3}$ the speed of sound in the plasma. The quantity $HR$ is typically determined at the percolation temperature $T_\mathrm{p}$, however, and provided that a large supercooling does not occur as in our numerical analysis, $T_\mathrm{p} \approx T_\mathrm{n}$ and both \cref{eq:fpeak,eq:HR} can be calculated at $T_\mathrm{n}$. Let us also introduce the fraction of the kinetic energy in the fluid to the total bubble energy as
\begin{equation}
	K = \dfrac{\kappa \alpha}{1 + \alpha}
	\label{eq:K}
\end{equation}
where fits to the efficiency factor $\kappa$ were taken from \cite{Espinosa:2010hh} and can be consulted in \cref{sec:App-A} for an easy reference. Another important quantity is the shock formation time-scale which quantifies the time that the source of GW lasted. Using \cite{Ellis:2019oqb,Caprini:2019egz} this can be written as
\begin{equation}
	H \tau_\mathrm{sh} = \dfrac{2}{\sqrt{3}} \dfrac{HR}{K^{1/2}}\,.
	\label{eq:Htau}
\end{equation}
If the source lasted less than the Hubble time, that is $H \tau_\mathrm{sh} < 1$, then the peak energy density today reads
\begin{equation}
	h^2 \Omega_\mathrm{GW}^\mathrm{peak} = 1.159 \times 10^{-7} \left(\dfrac{100}{g_\ast}\right)  \left(\dfrac{HR}{\sqrt{c_s}}\right)^2 K^{\tfrac{3}{2}}\,,
	\label{eq:Opeak1}
\end{equation}
while for the case of a source lasting approximately the Hubble time the amplitude of GW gets enhanced taking the form
\begin{equation}
h^2 \Omega_\mathrm{GW}^\mathrm{peak} = 1.159 \times 10^{-7} \left(\dfrac{100}{g_\ast}\right)  \left(\dfrac{HR}{c_s}\right)^2 K^{2}\,,
\label{eq:Opeak2}
\end{equation}
with the numerical factor on the r.h.s of both \cref{eq:Opeak1,eq:Opeak2} can be taken from \cite{Caprini:2019egz}. Finally, the GW spectrum for various frequencies $f$ can be taken by multiplying the peak amplitude by the spectral function and reads
\begin{equation}
	h^2 \Omega_\mathrm{GW} = h^2 \Omega_\mathrm{GW}^\mathrm{peak} \left(\dfrac{4}{7}\right)^{-\tfrac{7}{2}} \left(\dfrac{f}{f_\mathrm{peak}}\right)^3 \left[1 + \dfrac{3}{4} \left(\dfrac{f}{f_\mathrm{peak}}\right) \right]^{-\tfrac{7}{2}}\,.
	\label{eq:spectrum}
\end{equation}
Note that \cref{eq:Opeak1,eq:Opeak2,eq:spectrum} are valid for deflagrations with bubble wall velocities below the Chapman-Jouguet speed $v_\mathrm{b} < v_\mathrm{J} = c_s$ or for detonations with wall velocities above the Chapman-Jouguet speed $v_\mathrm{b} > v_\mathrm{J}$ with $v_\mathrm{J}$ given in \cref{eq:vJ}. In what follows we will study supersonic detonations with $v_\mathrm{b} > v_\mathrm{J}$.

\subsection{Properties of GWs spectra from separate phase transitions}
\label{Sect:GWs-properties}

Let us discuss now the basic characteristics of the GWs spectra focussing on separate weak and 
strong first-order phase transitions in the 2HDSM scenario. In the analysis below, we set up a generic
large scan at a computer cluster performed over the parameter space of the model with full one-loop $T$-dependence 
effective potential implemented in the \texttt{CosmoTransitions} package \cite{Wainwright:2011kj}.
We set the physical masses $M_{s_{1}}$, $M_{s_{2}}$ and $M_{s_{3}}$ as well as the quartic couplings $\lambda_2$, $\lambda_3$, $\lambda_s$, $\lambda_{s1}$ and $\lambda_{s2}$ as input parameters, using \cref{mass} to determine $\lambda_3^\prime$, $\mu_1^2$, $\mu_2^2$ and $\mu_s^2$. While the scalar masses are randomly generated in the linear interval $[50, 550]~\mathrm{GeV}$, the quartic couplings are logarithmically sampled within the range $\log_{10}(\lambda_i) \in [-3,1]$.
We find various possible phase transition patterns this way and for each transition we compute all its basic characteristics
needed for consistent evaluation of the produced GW spectra. For single-step transitions in particular we employ 
the stater of the art formalism in Ref.~\cite{Caprini:2019egz} for derivation of 
the associated GWs spectrum, $h^2 \Omega_{\rm GW}$. As was mentioned above, in the typical multi-scalar
scenarios such as the one considered in this work the collisions of bubble walls do not take part 
in the production of GWs unless unrealistic bubble runaway configurations with abnormally large $\alpha$ 
are concerned (for more details, see a discussion in Ref.~\cite{Hindmarsh:2017gnf,Ellis:2019oqb}). The lack of a yet solid knowledge about turbulence effects lead us to disregard its effect in GW production. In fact, it is believed that the dominant effect for both the peak frequency and amplitude, which are our key observables, comes from sound wave contributions. Therefore, the GW signal computed as the energy density per logarithmic frequency of the GW radiation solely considers the SW component generated by bubble expansion (see above). Since being produced at very early stages of the cosmological evolution, these signals get further effectively redshifted contributing to the stochastic GW background probed by a GW spectrometer.
\begin{figure}
\centering
\includegraphics[width=0.495\linewidth]{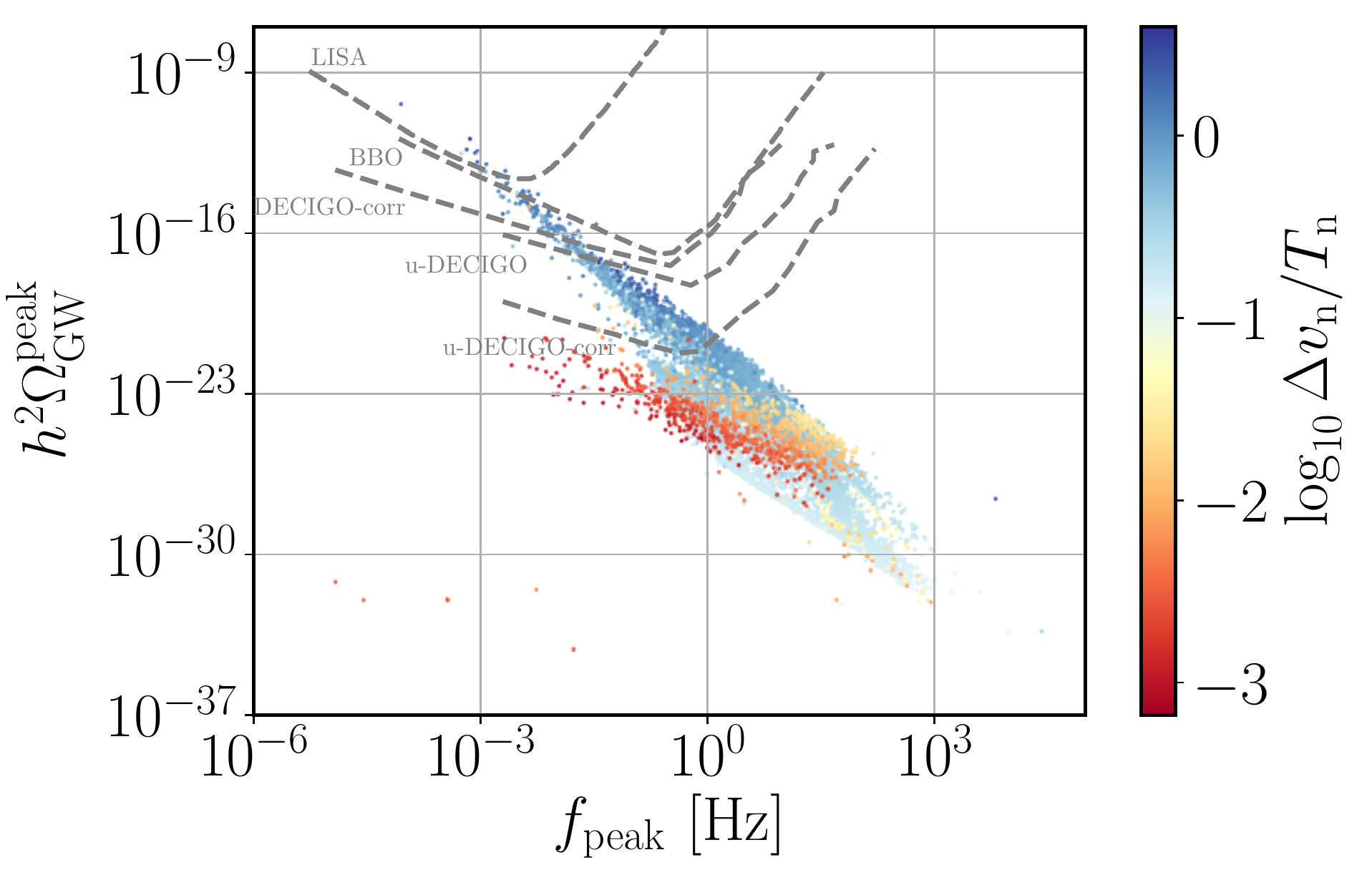}
\includegraphics[width=0.495\linewidth]{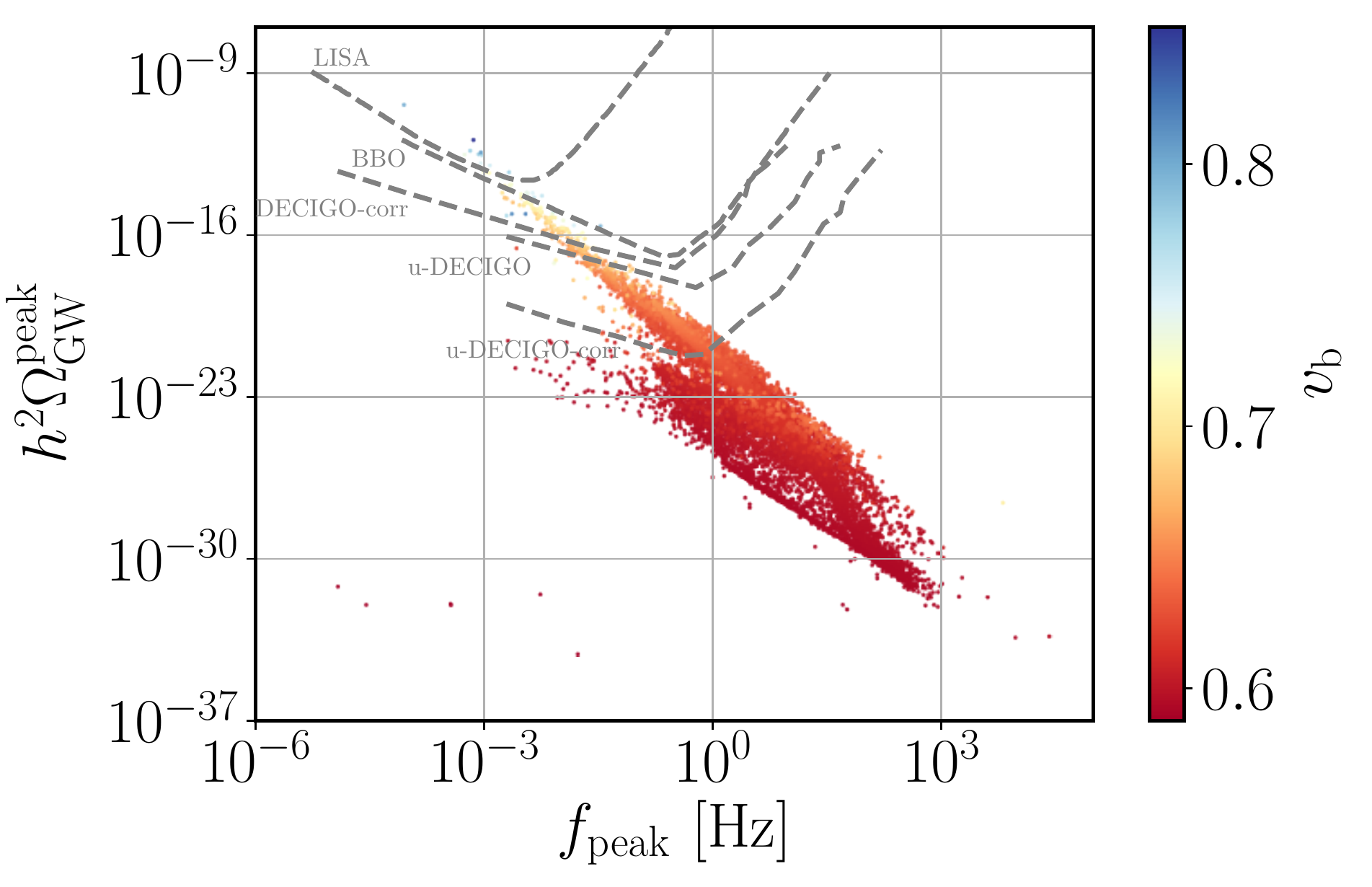}
\caption{Scatter plots showing the typical strength of the phase transitions $\Delta v_n/T_n$ entering Eq.~(\ref{strongPT_nucleation}) (left panel) 
and the characteristic bubble wall velocity $v_b$ (right panel) against the peak value of the corresponding net produced GW signal, 
$h^2\Omega_{\rm GW}^{\rm peak}$ , and its peak frequency, $f_{\rm peak}$, 
in the considered 2HDSM scenario. Here and below, dashed grey lines represent merely indicative sensitivities of the LISA \cite{Audley:2017drz} interferometer, 
as well as the proposed DECIGO \citep{Seto:2001qf,Kudoh:2005as,Kawamura:2011zz,Kuroyanagi:2014qaa} 
and BBO \citep{Crowder:2005nr,Corbin:2005ny} missions (see also Ref.~\citep{Buonanno:2004tp}). The power-law sensitivity curves 
for LISA were extracted from Ref.~\citep{Caprini:2015zlo} where we have taken the most optimistic approach and considered the old configuration N2A5M5L6 (see Tab.~1 and Fig.~3 of \citep{Caprini:2015zlo} for details). While the BBO sensitivity curve can be taken from \citep{Thrane:2013oya,Moore:2014lga}, those for the ultimate DECIGO, as well as for the DECIGO and ultimate-DECIGO 
with correlation analysis (denoted as ``DECIGO-corr'' and ``ultimate-DECIGO-corr'') can be found e.g.~Ref.~\cite{Nakayama:2009ce} that makes use of 
the sensitivity and signal-to-noise ratio results of Ref.~\cite{Kudoh:2005as}.
}
\label{fig:GW_vnTn-vb}
\end{figure}

For simplicity, we adopt ${\cal H}_1=\{v_h,0,0\}$ to be the stable vacuum state asymptotically at $T=0$ and 
ensure its stability by imposing the positivity of the scalar mass spectrum (\ref{mass}), the BFB conditions (\ref{BFB}) 
and the perturbativity constraints on quartic self-interactions, $|\lambda_i|<10$. Note, we do not restrict 
ourselves to any particular set of initial states and the phase transition patterns {\it \'a la} those 
in Eqs.~\eqref{Eq:pat1} and \eqref{Eq:pat2} discussed in the previous section. So, the results presented 
here and below are generic enough to represent all potentially interesting scenarios in the 2HDSM 
from the GWs phenomenology point of view.

In what follows, we show the scatter plots where each point represents a particular phase transition 
found for a given parameter space point in the 2HDSM scenario generated by our
simulation. For each such phase transition, we have collected all potentially relevant information 
about its characteristics and, most importantly, have evaluated the key quantities needed for
building the GWs spectrum produced in such a transition. In all the scatter plots below, the same phase 
transition points are shown focussing on their different characteristics.

From the phenomenological perspective, the most relevant quantity is the peak value of the GW power spectrum, 
denoted as $h^2\Omega_{\rm GW}^{\rm peak}$ (see \cref{eq:Opeak1,eq:Opeak2}), as well as the corresponding
peak frequency $f_{\rm peak}$ (see \cref{eq:fpeak}). Despite that the largest density and the amount of points found in our
simulation emerge below the projected sensitivities of near-future and proposed interferometers, a subset of such
transitions are at the reach of BBO and even LISA whose data therefore may set potentially relevant constraints on the 2HDSM model parameter space.

Particularly, in Fig.~\ref{fig:GW_vnTn-vb}, we show the distribution of all the phase transition points found in our
simulation together with the sensitivity curves of both planned and proposed GW interferometers, where the colour scheme represents the order parameter $\Delta v_n/T_n$ in logarithmic units (left panel), as well
as the bubble wall velocity $v_b$ that maximizes the peak amplitude such that $v_\mathrm{b} > v_\mathrm{J}$ (right panel). For the phase transition points we have collected 
here, there is a mild correlation between $h^2\Omega_{\rm GW}^{\rm peak}$ and $f_{\rm peak}$ values such that
most of the points are accumulated along a bend stretched between the upper left and lower right corner
of the figure. Often, larger GW amplitudes generally prefer smaller frequencies, with some small islands 
of points somewhat deviating from this trend. Remarkably, in the multi-scalar model under consideration a relatively large portion of the generated set of blue points, corresponding mostly to the FOPTs with large $\Delta v_n/T_n\sim 1$ ratio thus potentially relevant for electroweak baryogenesis,
can be probed by future (or proposed) GW interferometers.
\begin{figure}
\centering
\includegraphics[width=0.495\linewidth]{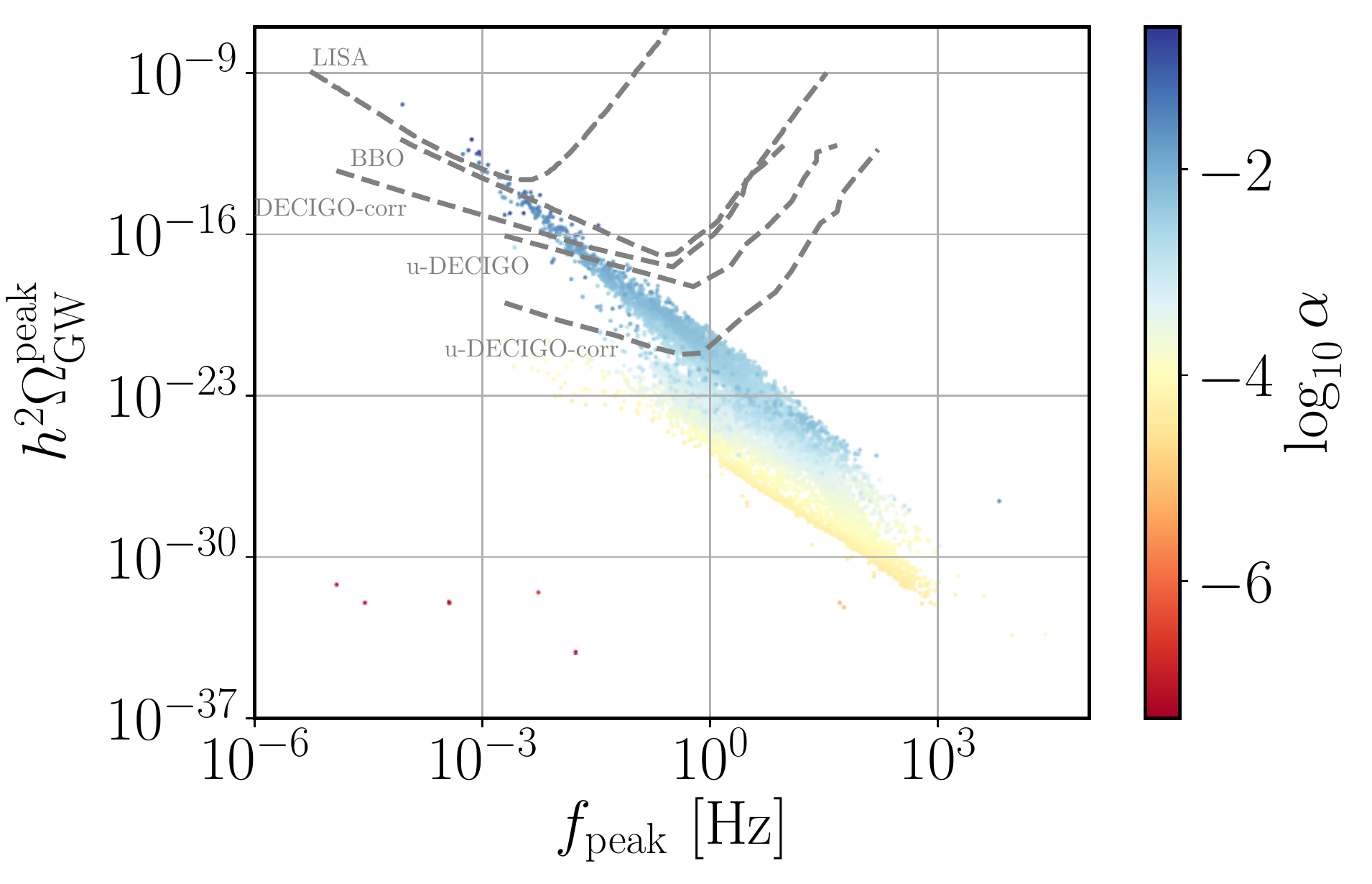}
\includegraphics[width=0.495\linewidth]{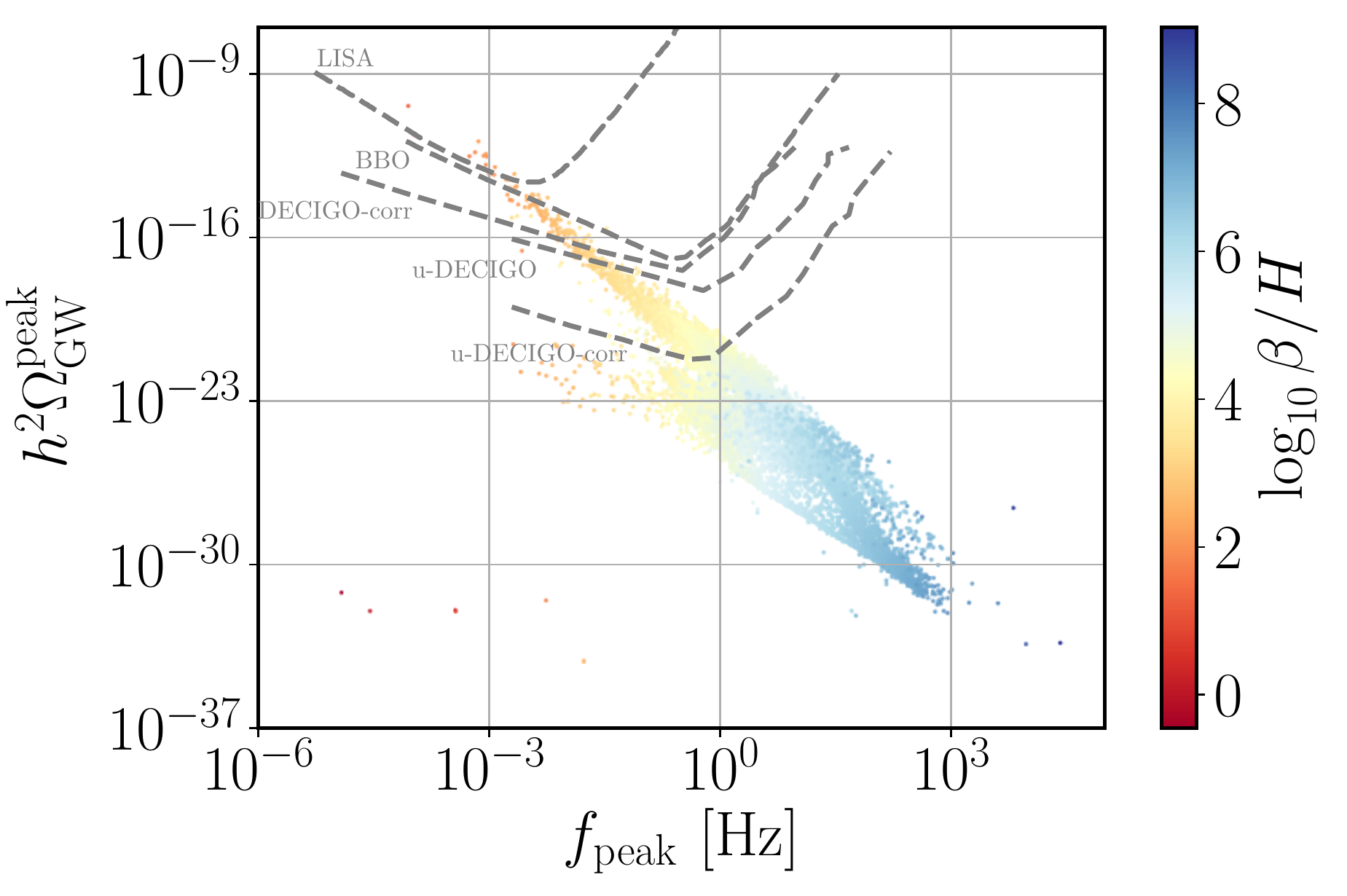}
\caption{Scatter plots showing the latent heat of the phase transition $\alpha$ given by Eq.~\eqref{alpha} 
(left panel) and the inverse time-scale of the phase transition in units of the Hubble parameter $H$, 
$\beta/H$, found in Eq.~\eqref{betaH} (right panel). Both quantities are given on the logarithmic scale 
against the peak value of the corresponding net produced GW signal, $h^2\Omega_{\rm GW}^{\rm peak}$, 
and its peak frequency, $f_{\rm peak}$, in the 2HDSM scenario under consideration.
}
\label{fig:GW_alphabetaH}
\end{figure}

We notice that for the majority of FOPT points there is an apparent correlation also between the ratio $\Delta v_n/T_n$ 
and the magnitude of the corresponding peak in the GW power spectrum such that larger $\Delta v_n/T_n$ often correspond to larger values of $h^2\Omega_{\rm GW}^{\rm peak}$ and somewhat 
smaller frequencies, although for the latter such a correlation is minor. Quite a few points that exhibit a large $\Delta v_n/T_n\sim 1$ ratio have been found potentially within LISA sensitivity domain, and this trend is clearly correlated with the strength of the transition trend shown in Fig.~\ref{fig:GW_alphabetaH} (left panel).
Often but not always, such transitions are strong first-order ones already at tree level. However, not that the correlation between the strength of the transition and the order parameter is not always universal. Namely, roughly in the middle of the plot we discover a sparse but rather populated family of red points that overlap with
many blue and yellow points and also stretch towards somewhat lower frequencies. This means, quite remarkably,
that a few observable GW signatures within the proposed ``u-DECIGO-corr'' sensitivity domain may also arise even from transitions, with $\Delta v_n/T_n$ ratio having quite low $0.01-0.1$ values. This is the reason why the criterion suggested
in Eq.~(\ref{strongPT_nucleation}) does not unambiguously and uniformly represents a good PT strength criterion since in some 
cases there is a strong anti-correlation of the $\Delta v_n/T_n$ ratio value with the GW peak-amplitude. On the contrary, the $\alpha$-parameter does indeed offer a reliable criterion to classify the strength of the PT. We consider a few such points among our benchmark scenarios below.

The bubble wall velocity is chosen in such a way that it maximizes the GW peak amplitude and takes typical values ranging between 0.6 and 0.9. We have also required it to be above the Chapman-Jouguet velocity given in \cref{eq:vJ} so that the formalism presented above and recently developed in \cite{Caprini:2019egz} applies.
The correlation between the peak value of the GWs spectrum and $v_b$ is rather uniform, such that
$h^2\Omega_{\rm GW}^{\rm peak}$ gradually increases with the growth of the wall velocity, while 
the peak frequency has a tendency to decrease with $v_b$. The characteristic values of $v_b$ that 
correspond to potentially observable GWs signals by LISA and BBO experiments lie beyond 0.8-0.9.

In Fig.~\ref{fig:GW_alphabetaH}, for each of the phase transition points shown in Fig.~\ref{fig:GW_vnTn-vb} 
we illustrate the strength $\alpha$ that is given by Eq.~(\ref{alpha}) (left panel) and the inverse 
time-scale of the phase transition in units of the Hubble parameter $H$, $\beta/H$, found 
in Eq.~(\ref{betaH}) (right panel) -- both are shown in logarithmic scale in the colour bar. Again, 
a clear correlation between the magnitude of the GW peak-amplitude, $h^2\Omega_{\rm GW}^{\rm peak}$,
the corresponding peak-frequency, $f_{\rm peak}$, with respect to both $\alpha$ and $\beta/H$ is observed. 
Quite expectedly, the stronger phase transitions with larger $v_b$ and $h^2\Omega_{\rm GW}^{\rm peak}$ 
generally have smaller $\beta/H$ values (hence, release larger amounts 
of heat and last longer on the time scale of the Universe evolution). Note that we find a few low-frequency and low amplitude points with $\alpha \lesssim 10^{-6}$ which are likely very weak cross-overs where perturbative analysis is less reliable.
\begin{figure}
\centering
\includegraphics[width=0.495\linewidth]{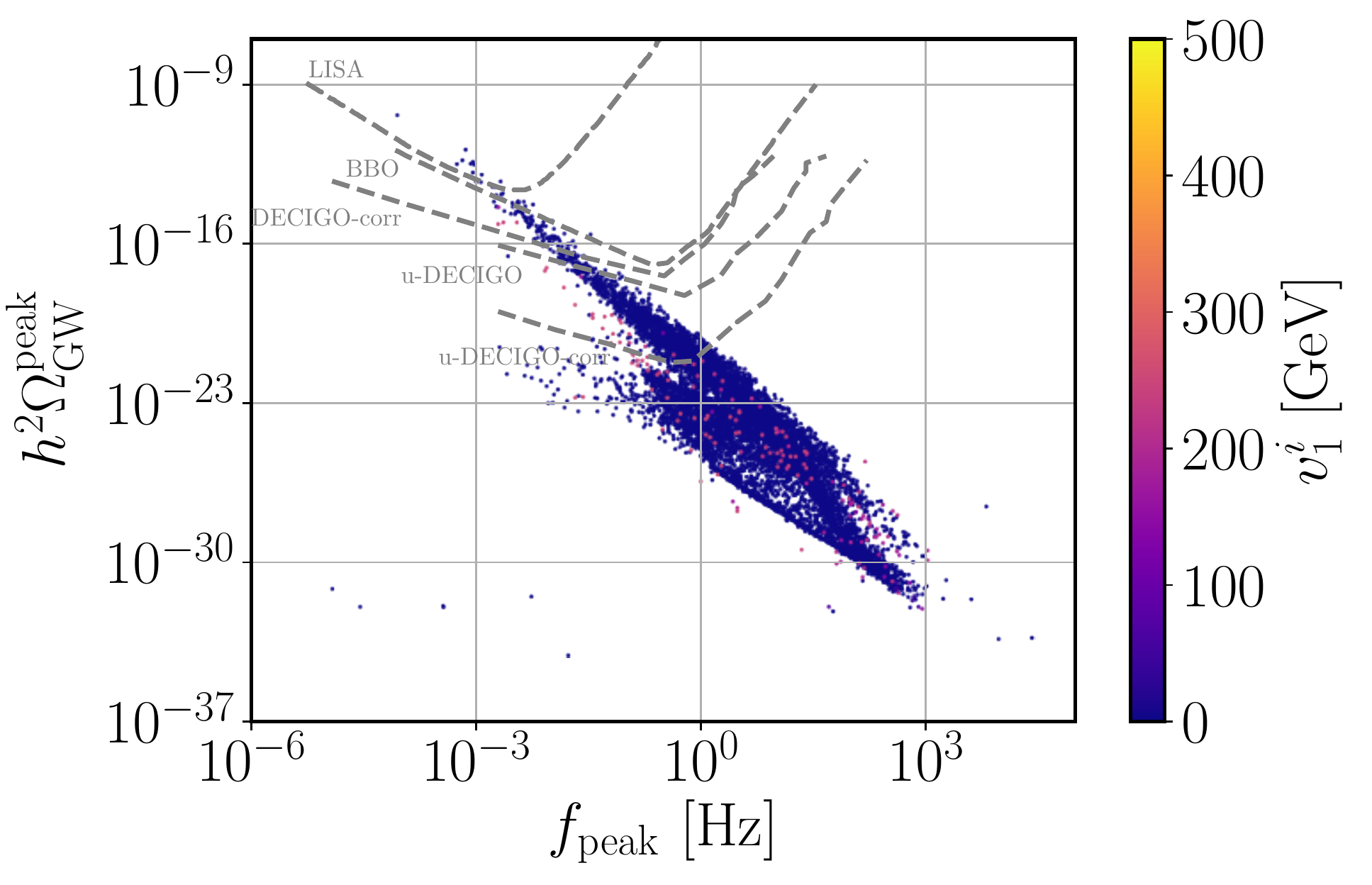}
\includegraphics[width=0.495\linewidth]{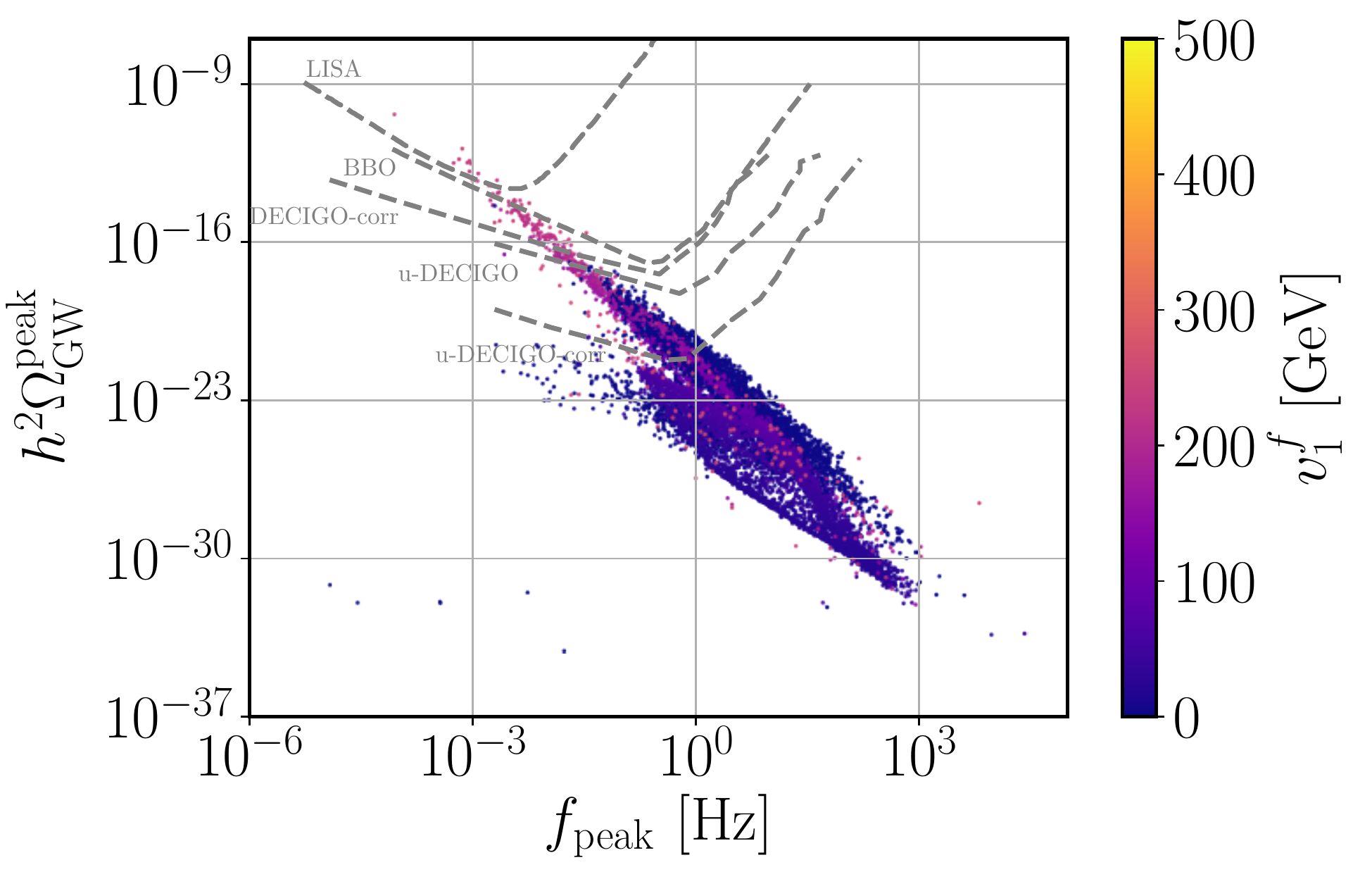}
\includegraphics[width=0.495\linewidth]{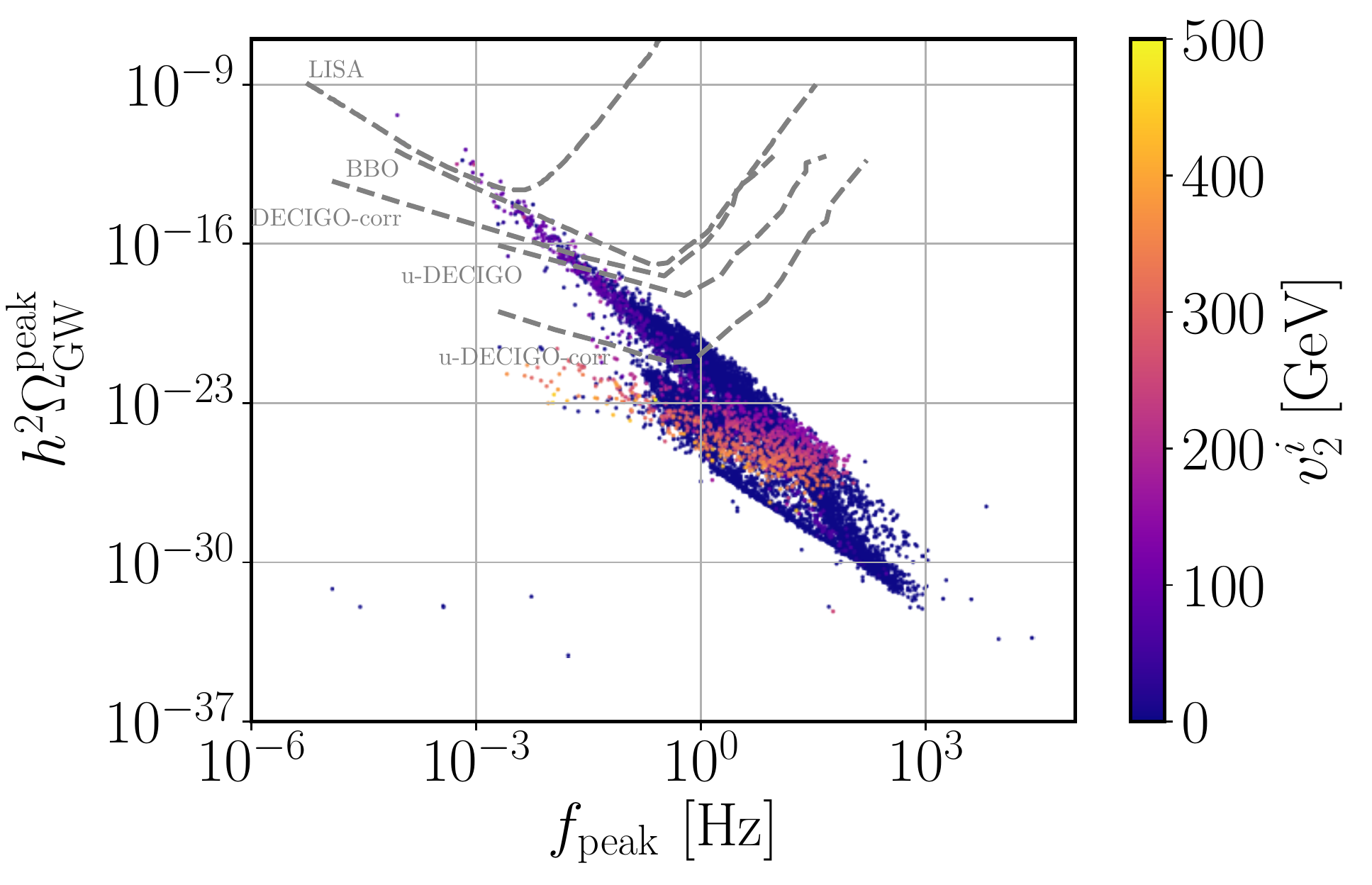}
\includegraphics[width=0.495\linewidth]{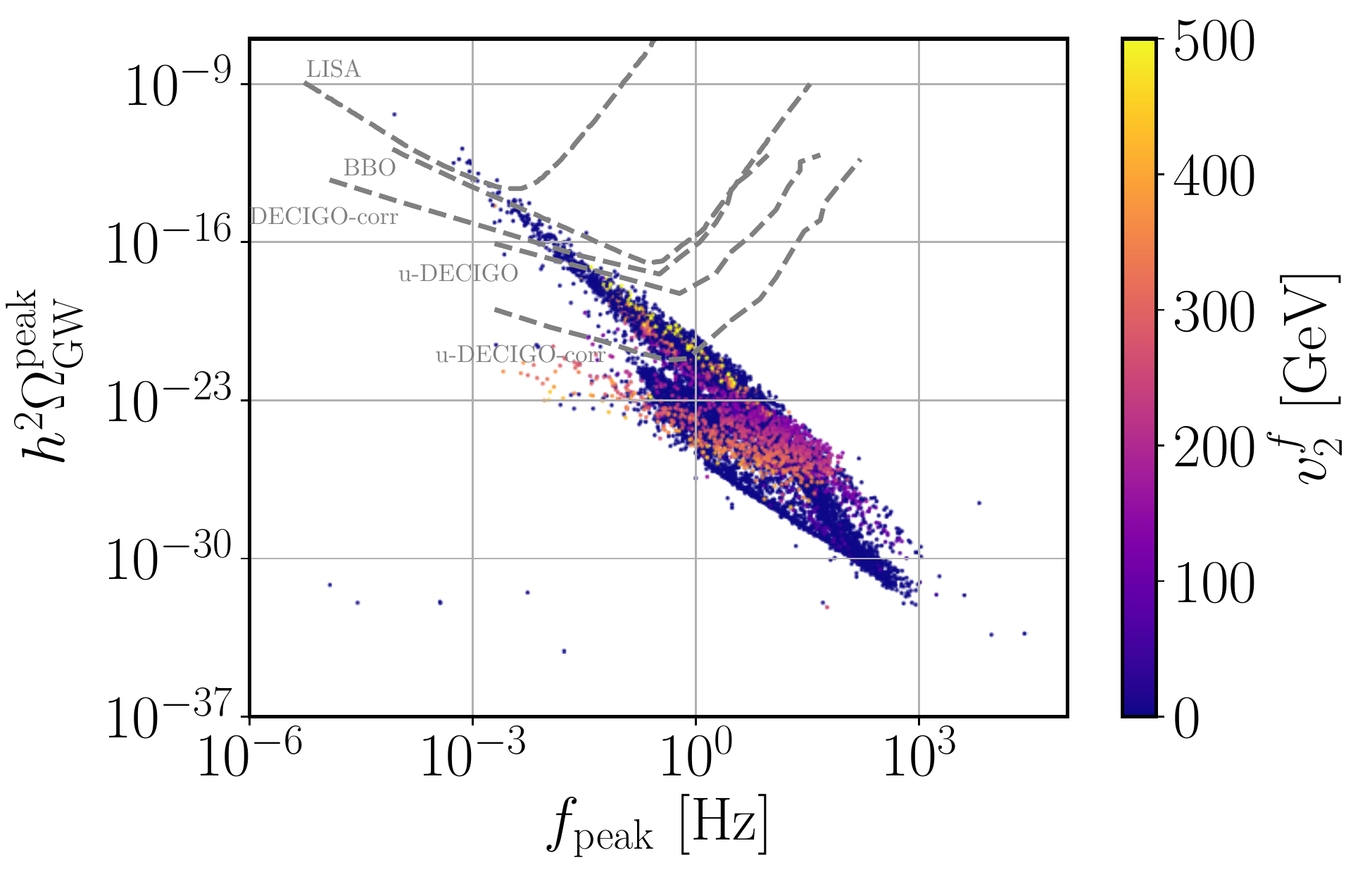}
\includegraphics[width=0.495\linewidth]{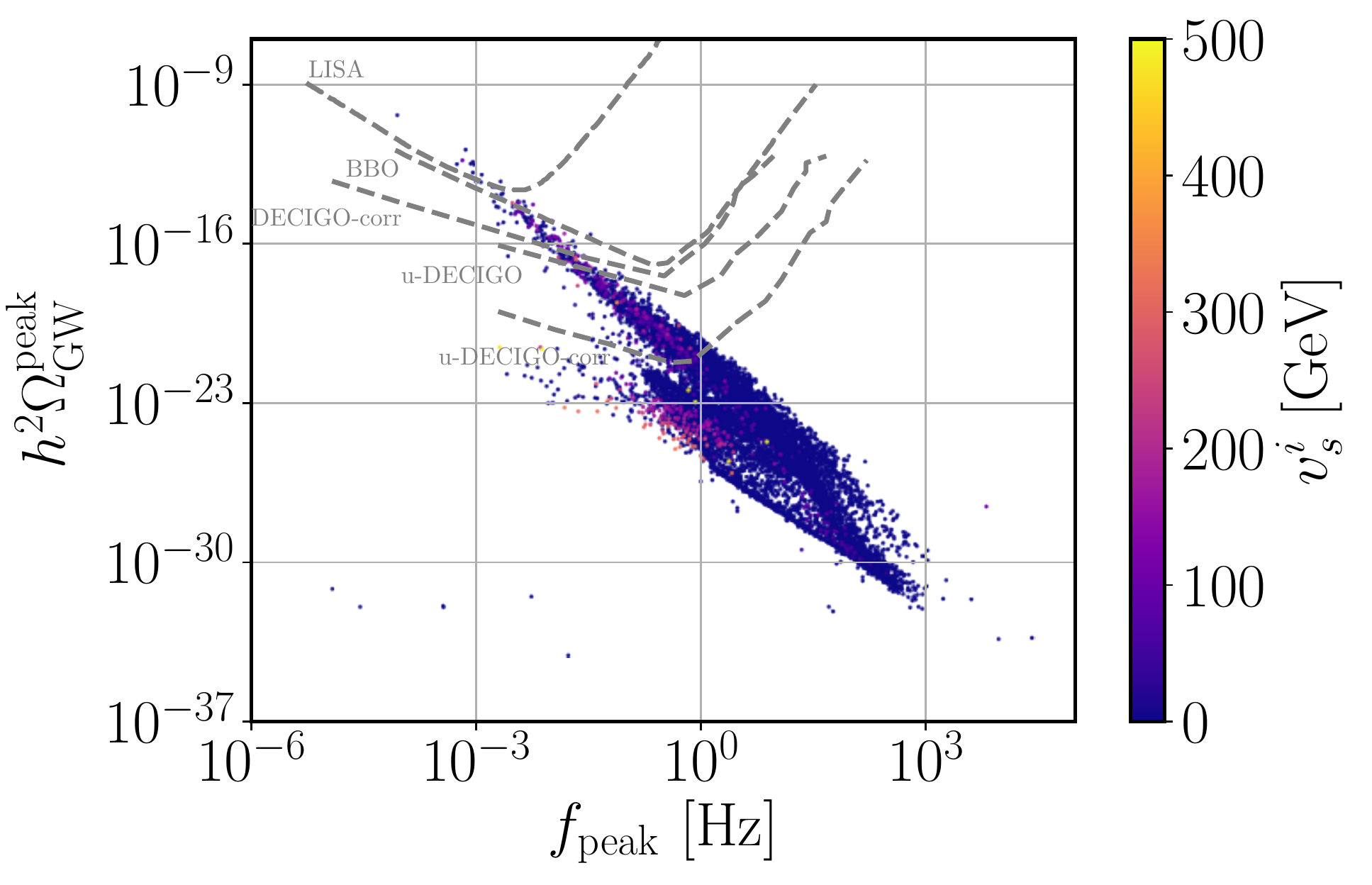}
\includegraphics[width=0.495\linewidth]{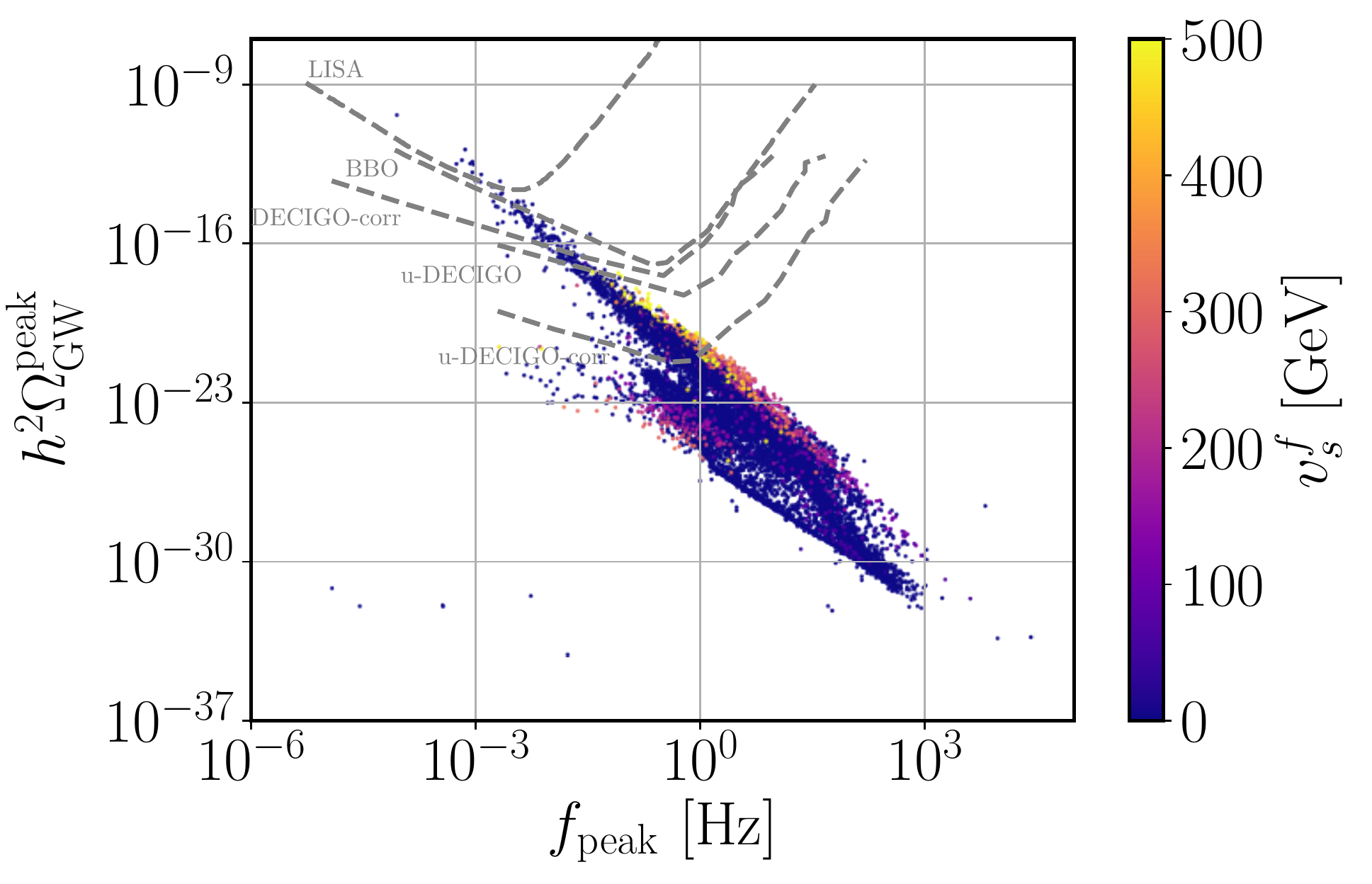}
\caption{Scatter plots showing the VEVs of the scalar fields, namely, for Higgs doublet $H_1$ (top panels), 
Higgs doublet $H_2$ (middle panels) and complex singlet $S$ (bottom panels), computed just before (left panels) 
and after (right panels) the corresponding phase transition, i.e. at $T^i= T_n - dT$ and $T^f= T_n + dT$ respectively. These quantities are provided 
together with the peak value of the corresponding net produced GW signal, $h^2\Omega_{\rm GW}^{\rm peak}$, 
and its peak frequency, $f_{\rm peak}$, in the considered 2HDSM scenario.
}
\label{fig:GW_VEVs}
\end{figure}

In Fig.~\ref{fig:GW_VEVs}, we present the VEVs of the scalar fields at finite temperatures corresponding to the classical field configurations just before $(v^i_1,v^i_2,v^i_s)$ (left panels) and after $(v^f_1,v^f_2,v^f_s)$ (right panels) the corresponding phase transition for each given point generated by our simulation. Such plots enable us to investigate the phase structure of the $T$-dependent vacuum.
Despite that some points overlay on top of each other, we clearly see some tendencies that are generally seen for large domains in each panel. For example, we notice a rather unique trend with a few very strong first-order transitions producing GWs signals in the LISA domain, where the initial phase corresponds to $v^i_1,v^i_s$ being either zero or small while $v^i_2$ can be as large as $\mathcal{O}(200~\mathrm{GeV})$, and the final phase contains $v^f_1 \sim 246$ GeV, while $v^f_2,v^f_s$ become or remain to be small. It is interesting to note that the red island in Fig.~\ref{fig:GW_vnTn-vb} (left panel) that overlaps with the blue continuous trend corresponds to scenarios where $v_2$ and $v_s$ can be large before and after the PT, thus contributing to the order parameter in \eqref{strongPT_nucleation}. Note, due to a specific structure of interactions and $\mathrm{U}\left(1\right)_\mathrm{F}$ charges in the 2HDSM scenario under consideration, the phase structure of $H_1$ and $H_2$ fields look somewhat different.

\subsection{GWs spectra from sequential phase transitions}
\label{Sect:GWs-sequential}

In our numerical analysis, the nucleation temperatures for two sequential phase transitions (a) and (b) satisfy 
$T_n^{(a)} - T_n^{(b)} > \Delta T\sim $ 10 GeV, such that bubbles nucleation in the transition (b) starts only after the bubbles 
of the transition (a) completely percolate. In this typical case, the corresponding first-order phase transitions 
are well-separated and occur at very different time scales such that the well known formalism of 
Ref.~\cite{Caprini:2019egz} for derivation of the GWs spectrum, 
$h^2 \Omega_{\rm GW}$, emerging from single-step transitions is justified. For successive well-separated 
transitions like the ones discussed here the net GW energy density is just the mere superposition of the corresponding 
contributions emerging from the single-step transitions yielding well-separated (in frequency) GWs signals, or peaks 
in the GW spectrum. Even though it is quite obvious that such a superposition should naturally lead to the multi-peaked 
signatures in the power spectrum of GWs, an explicit calculation in a particularly simple extension 
of the Higgs sector that adopts, at least, two such transitions is lacking the literature. 
\begin{figure}
\centering
\includegraphics[width=0.495\linewidth]{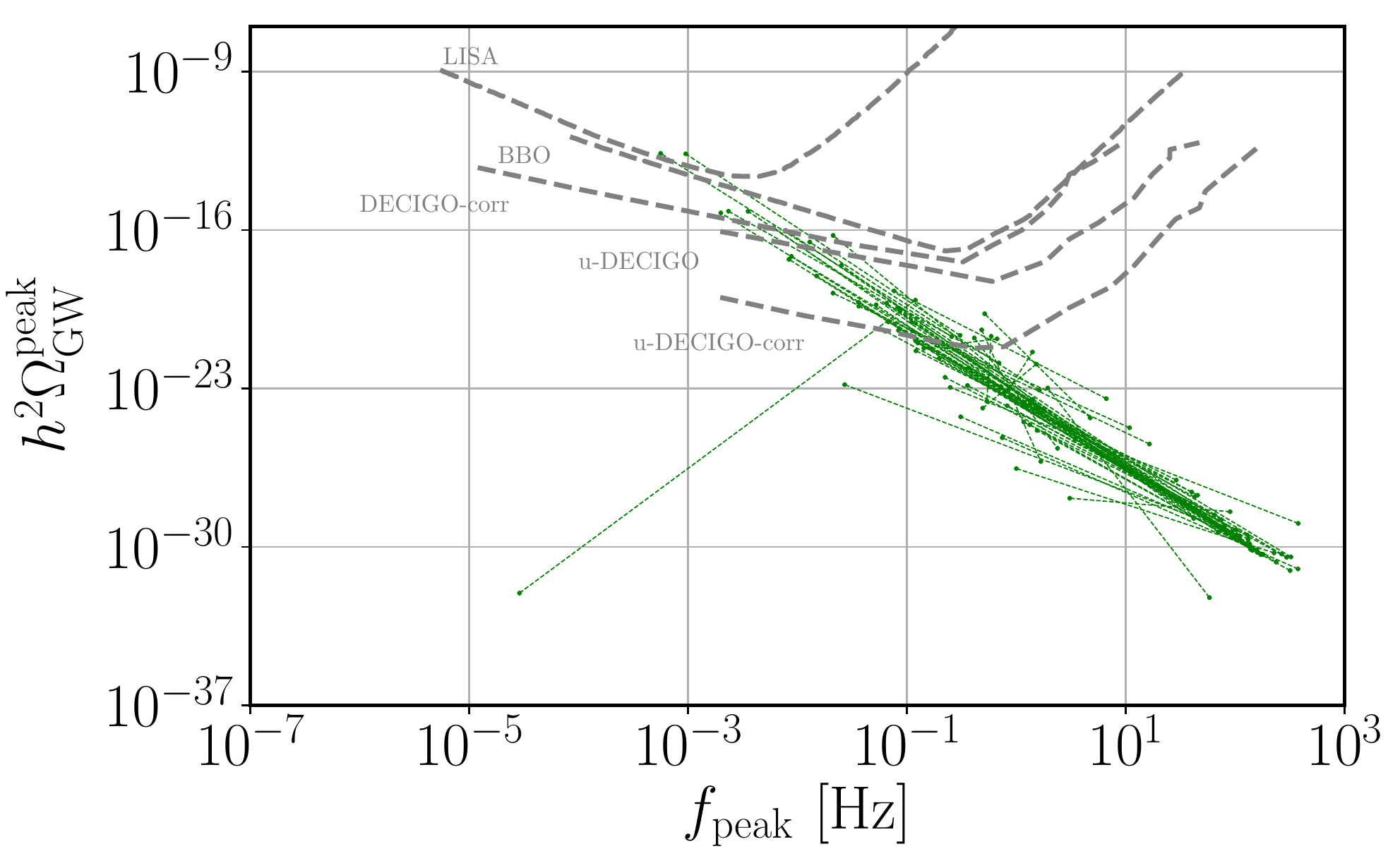}
\includegraphics[width=0.495\linewidth]{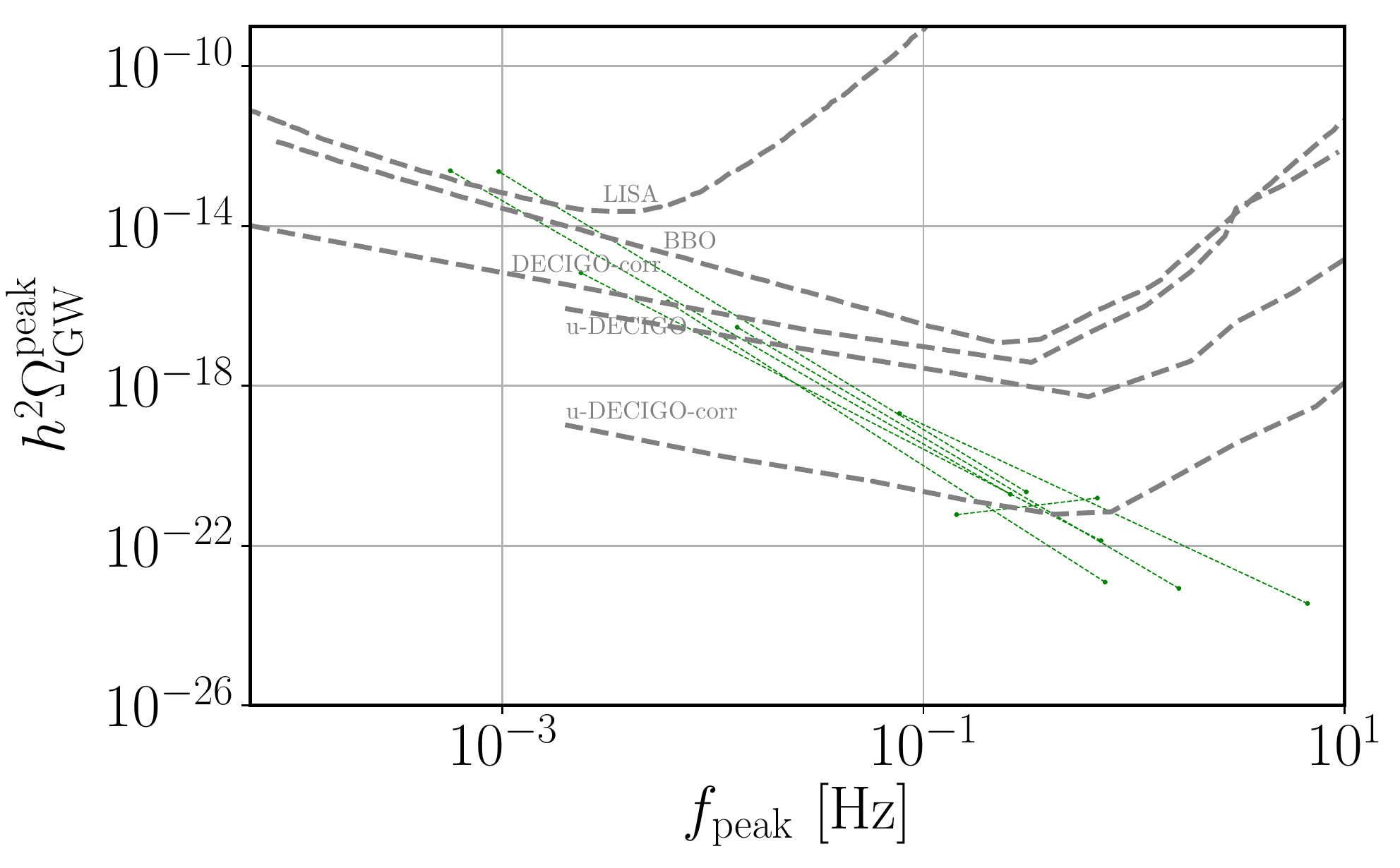}
\caption{An illustration of the double-peak GW signals where for each signal the two subsequent peak values generated 
by sequential first-order phase transitions are connected by a single line. In the left panel, we draw all the 
double-peak configurations found in our numerical scan such that the distribution of lines represent typical 
magnitudes and peak-frequencies for such signals. In the right panel, we show a selection of seven benchmark points with
potentially observable double-peak configurations which are identified as $id=1\dots 7$ as described in detail 
in Tables~\ref{tab:properties_GW} and \ref{tab:points}.
}
\label{fig:GW_Npeaks}
\end{figure}

For some of the parameter space points found in our numerical scan, we have identified up to two sequential 
phase transitions, with rather distinct characteristics. We extracted and presented in Fig.~\ref{fig:GW_Npeaks} (left panel)
all the double-peak GW spectra configurations by connecting two subsequent peaks by a straight line for each such
configuration. In other words, each line corresponds to a single double-peak GW spectrum generated by two sequential 
first order phase transitions found for a given parameter space in the 2HDSM. Despite of a few outliers, we notice that such double-peak configurations accumulate the lines approximately stretched between upper left and lower right conner connecting one big potentially visible GW signal and one much smaller peak. A few such scenarios 
with both peaks not far from a potentially observable domain of signals are isolated and shown in \cref{fig:GW_Npeaks} (right panel). 
\begin{table}
\centering
\resizebox{\columnwidth}{!}{%
\begin{tabular}{|c|c|ccccc|cccccc|cc|c|} \hline
$id$ & PT  &  $T_n$ &  $\Delta v_n$ & $\alpha$ & $\beta/H$ &  $v_b$  & $v^i_1$ & $v^f_1$ & $v^i_2$ & $v^f_2$ 
    & $v^i_s$ & $v^f_s$ & $f_{\rm peak}$ & $h^2\Omega_{\rm GW}^{\rm peak}$ & Order \\ \hline
1 & (a) & 196 & 110 & 7.5$\cdot 10^{-4}$ & 5.7$\cdot 10^{4}$ & 0.61  & 0      & 0       & 0       & 110    & 0     & 0 & 1.6 & 8.5$\cdot 10^{-24}$ & $O(2)$ \\
   & (b) & 172   & 36 & 8.0$\cdot 10^{-3}$ & 555  & 0.66  & 0     & 193        & 157     & 0   & 0  & 0 & 0.01 & 2.9$\cdot 10^{-17}$  & $O(2)$ \\ \hline
2 & (a) & 80 & 33 & 2.4$\cdot 10^{-3}$ & 6.1$\cdot 10^{4}$ & 0.63 & 0        & 87      & 86     & 81  & 0     & 0   & 0.7                       & 1.3$\cdot 10^{-22}$  & $O(2)$ \\
   & (b) & 6   & 5 & 0.14                          & 3.4$\cdot 10^{3}$ & 0.81 & 240    & 246   & 23     & 0     & 0    & 0   & 2$\cdot 10^{-3}$ & 6.7$\cdot 10^{-16}$  & $O(2)$ \\ \hline
3 & (a) & 194 & 175 & 1.9$\cdot 10^{-3}$ & 1.1$\cdot 10^{4}$ & 0.62 & 0        & 0       & 0        & 175  & 0   & 0   & 0.3                     & 2.2$\cdot 10^{-21}$  & $O(2)$ \\
   & (b) & 86   & 9 & 0.08                          & 96                           & 0.77 & 0        & 240   & 231    & 0      & 0   & 0   & 9.6$\cdot 10^{-4}$ & 2.3$\cdot 10^{-13}$  & $O(1)$ \\ \hline
4 & (a) & 335 & 164 & 1.1$\cdot 10^{-3}$ & 1.4$\cdot 10^{5}$  & 0.61  & 0        & 0   & 0        & 0     & 0    & 164  & 6.7 & 3.6$\cdot 10^{-24}$  & $O(2)$ \\
   & (b) & 48   & 38 & 0.013 & 1.2$\cdot 10^{4}$  & 0.67  & 0  & 0   & 38        & 0     & 349   & 349  & 7.7$\cdot 10^{-2}$ & 2.0$\cdot 10^{-19}$  & $O(2)$ \\ \hline
5 & (a) & 164 & 158 & 1.8$\cdot 10^{-3}$ & 1.1$\cdot 10^{4}$  & 0.62  & 0        & 0       & 0        & 158  & 0    & 0  & 0.3                     & 1.9$\cdot 10^{-21}$  & $O(2)$ \\
   & (b) & 91   & 28 & 0.047 & 51                           &  0.74  & 0        & 235   & 207    & 0      & 0    & 0  & 5.6$\cdot 10^{-4}$ & 2.4$\cdot 10^{-13}$  & $O(1)$ \\ \hline
6 & (a) & 136 & 85    & 6.0$\cdot 10^{-4}$ & 3.6$\cdot 10^{4}$  & 0.61  & 0      & 0       & 0        & 85   & 0     & 0  & 0.73                    & 1.2$\cdot 10^{-23}$  & $O(2)$ \\
   & (b) & 121   & 82 & 0.001 & 373                          &  0.66  & 0      & 198   & 116     & 0      & 0   & 0 & 6.1$\cdot 10^{-3}$& 1.3$\cdot 10^{-16}$  & $O(1)$ \\ \hline
7 & (a) & 166 & 153 & 3.5$\cdot 10^{-3}$ & 2.9$\cdot 10^{4}$   & 0.63  & 0         & 153   & 0      & 0      & 0     & 0  & 0.67 & 1.6$\cdot 10^{-21}$  & $O(2)$ \\
   & (b) & 24   & 15 & 3.3$\cdot 10^{-3}$ & 4.2$\cdot 10^{4}$  & 0.63  & 246    & 246   & 0      & 0      & 15   & 0  & 0.14 & 6.0$\cdot 10^{-22}$  & $O(2)$ \\ \hline
\end{tabular}%
}
\caption{Properties of a few selected sequential (double) transitions whose peaks appear near the sensitivity ranges of proposed and planned measurements. Here, the nucleation temperature, $T_n$, the difference between the order parameter $v(T)$ values computed 
before and after a given phase transition, $\Delta v_n$ (see Eq.~(\ref{strongPT_nucleation})), 
the scalar VEVs before $v^i_\alpha$ and after $v^f_\alpha$ the respective phase transition are given in units of GeV, while 
the peak-frequency, $f_{\rm peak}$, is given in Hz. The index $id=1\dots 7$ denotes distinct parameter space 
points of the 2HDSM extension of the SM under consideration specified in Table~\ref{tab:points}. For each such parameter 
space point, two sequential transitions (ordered in $T_n$) have been found and are denoted as (a) and (b) such 
that $T_n^{(a)}>T_n^{(b)}$. The last column indicates the order of the phase transition at tree-level according to the generic classifications in \cite{Vieu:2018nfq}. Such points can be further considered as benchmarks for further explorations 
at GW interferometers.
}
\label{tab:properties_GW}
\end{table}

Several benchmark examples of double phase transitions illustrated in \cref{fig:GW_Npeaks} (right panel) and labelled 
by $id=1\dots 7$ are also presented in \cref{tab:properties_GW,tab:points} providing a detailed information of their properties. 
The corresponding 2HDSM model parameters for each such double transition are given in \cref{tab:points}. Among the potentially visible GW signals, 
we choose three particular representative  benchmarks for which two of them have the highest GW peak-amplitude potentially in the range of LISA 
and another one with similar peak amplitudes only accessible at proposed GW interferometers.
For the latter three transitions corresponding to $id=3,5$ and $7$ in \cref{tab:properties_GW,tab:points}
we plot in \cref{fig:GW_visible} their full GW spectra to also show the typical shape of such double-peak transitions.
Indeed, we notice from this figure that for the blue and green curves, although the peaks tend to be well-separated in frequency, the second peak is rather small and a significant detector resolution or advanced experimental techniques would be required for its reconstruction. Of course, the tail of the first biggest peak gets modified by the presence of the second one potentially inducing an observable
difference with respect to typical single-peak configurations. The larger frequencies however become challenging to observe 
\footnote{Very high GW frequencies have also been reached earlier in Ref.~\cite{Wan:2018udw}.}.
On the other hand, the red curve represents an interesting scenario where both peaks are relatively close in both amplitude and frequency where instead of two pronounced peaks the spectral shape approaches to a plateau at its maximum. This type of GW spectrum can be seen as a representative example of scenarios where the total energy budget is evenly distributed between both transitions as can be seen from \cref{tab:properties_GW}. However, the price to pay for such scenarios is a significant reduction on the amplitude of both peaks. While ground-based spectrometers such as LIGO or VIRGO can probe larger frequencies, they are not so sensitive to the typical range of small amplitudes corresponding to the second peaks in most of our generated double-peak configurations.

Recall that in our scan we have chosen bubble wall velocities that maximize the peak amplitude of each generated point. However, this is not necessarily the case and the effect of different wall velocities should be commented. In particular, if we select the blue curve in \cref{fig:GW_visible}, benchmark point $id=5$, and allow $v_\mathrm{J} < v_\mathrm{b} < 1$ the position of each individual peak would change according to \cref{fig:vb-peak}. The observed effect is generic for any other point in our scan where larger wall velocities reduce the conversion efficiency of vacuum energy into kinetic energy. In particular, the limit of large velocities suppress the amplitude of the higher peaks in both the blue and green curves below LISA reach whereas lower peaks end up hidden below the tail of the former. However, due to the distinct nature of (a) and (b) transitions (see discussion below) there is no reason for both of them generating equally large wall velocities. For instance, if the ``strong''-transition (higher peak) represents a bubble wall velocity larger than that of the ``weak''-transition (lower peak), then, the latter would become further resolved in comparison to what we see in \cref{fig:GW_visible}.
\begin{figure}
	\centering
	\includegraphics[width=0.7\linewidth]{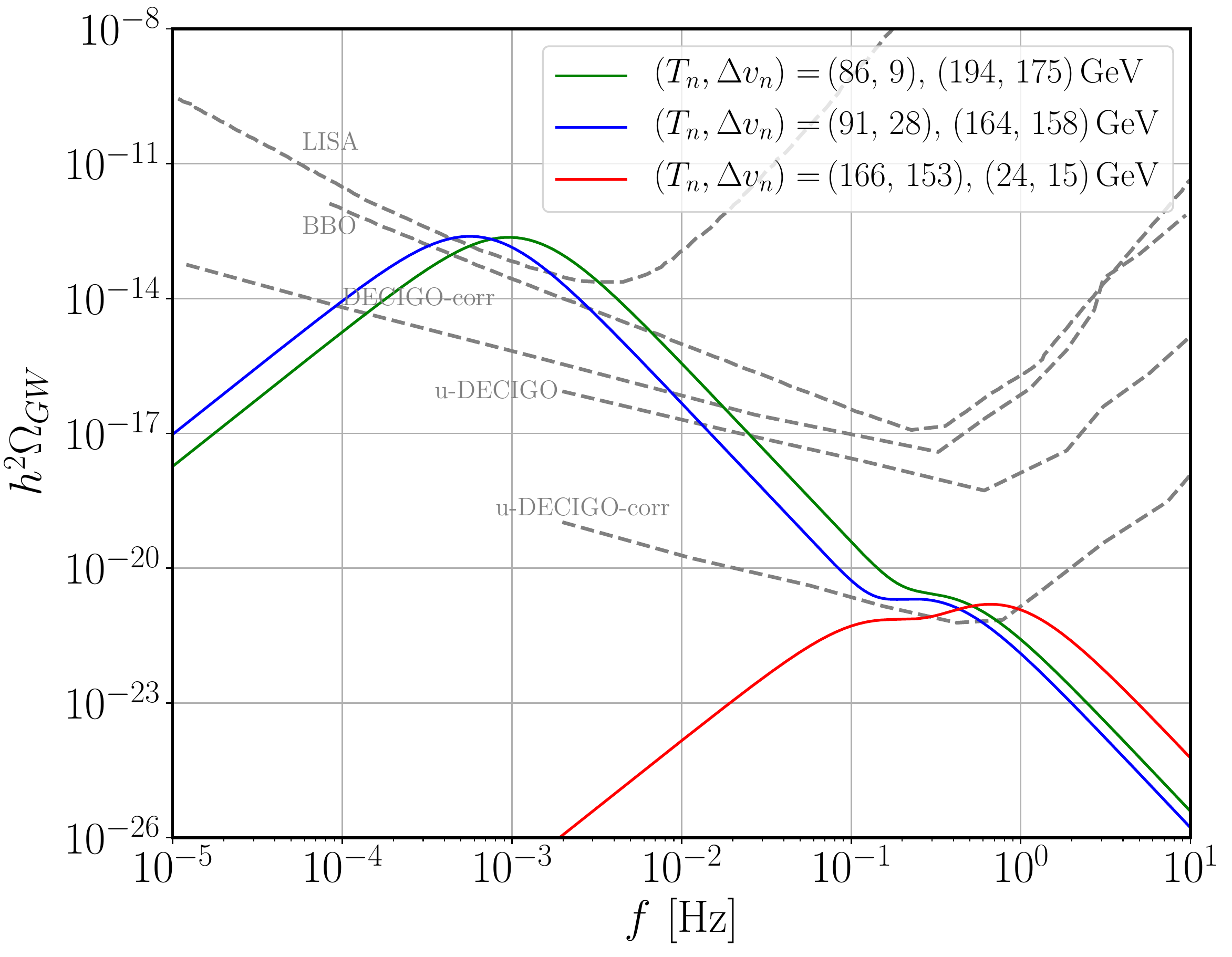}
	\caption{Three selected double-peak GW spectra, with the largest peaks in the sensitivity range 
		of the LISA interferometer. In Tables~\ref{tab:properties_GW} and \ref{tab:points}, these spectra correspond 
		to the benchmark points with $id=3,5$ and 7 (green, blue and red lines, respectively).
	}
	\label{fig:GW_visible}
\end{figure}
\begin{figure}
	\centering
	\includegraphics[width=0.9\linewidth]{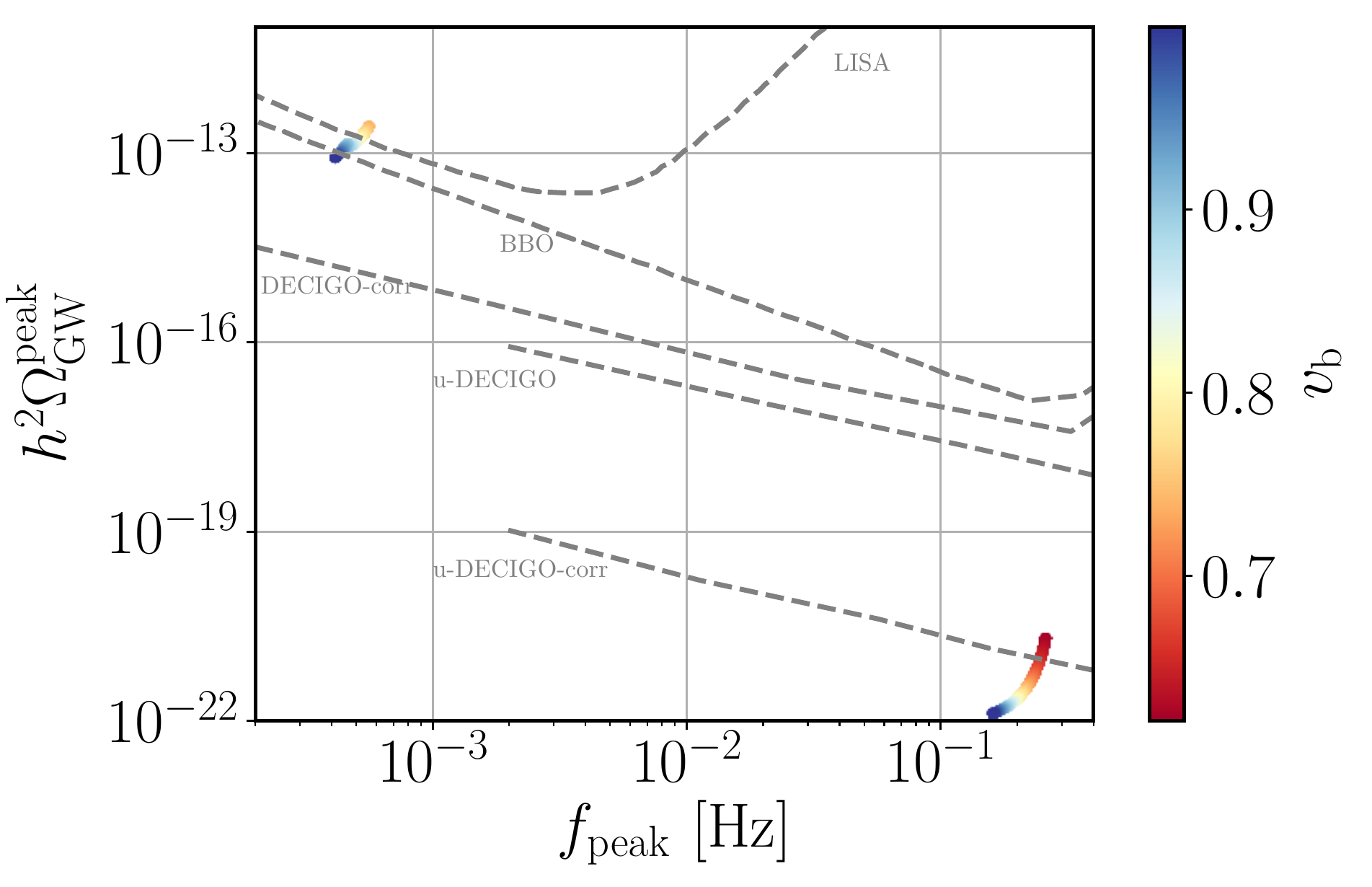}
	\caption{Peak positions for distinct wall velocities for the benchmark point $id=5$. The highest peak positions correspond to the peaks of the blue curve in \cref{fig:GW_visible}. }
	\label{fig:vb-peak}
\end{figure}

\begin{table}
\centering
\begin{tabular}{|c|ccccccccc|} \hline
$id$   &  $M_{s1}$ & $M_{s2}$ & $M_{s3}$ & $\lambda_2$ & $\lambda_3$ 
      & $\lambda_s$ & $\lambda_{s1}$ & $\lambda_{s2}$ & $\lambda'_{3}$ \\ \hline
1    &  66  & 535 & 482 & 7.4      & 0.1  & 0.4  & 2.2     & 6.6 &  0.003  \\ \hline
2    &  376 & 121  & 307 & 9.0      & 6.6  & 0.02    & 0.5     & 0.3     & -4.2  \\ \hline
3    &  511  & 98   & 122  & 0.03    & 8.8  & 0.005  & 0.1     & 0.01    & -8.3  \\ \hline
4    &  93   & 239 & 421    & 7.2  & 0.7 & 0.06     & 7.6 & 0.06 & 1.6  \\ \hline
5    &  444 & 115 & 347  & 0.1       & 6.7   & 0.03    & 0.06   & 0.03  & -6.1  \\ \hline
6    &  59   & 374 & 368 & 0.3       & 0.3   & 0.004 & 0.001 & 0.02   & 4.5  \\ \hline
7    &  105 & 475 & 74    & 0.002  & 0.09 & 9.3     & 0.2     & 0.01    & 7.1  \\ \hline
\end{tabular}
\caption{Specification of 2HDSM parameter space points denoted by an index ``id'' corresponding 
to the double phase transition benchmarks listed in Table~\ref{tab:properties_GW}. 
(The lightest Higgs boson mass is fixed to the observed value $m_h=125$ GeV).
}
\label{tab:points}
\end{table}

Let us discuss basic qualitative features of the selected benchmarks. As we mentioned earlier, 
some of these FOPTs are strong enough to produce potentially visible GW signatures at the proposed next-generation 
GW interferometers. Given very different nucleation temperatures, we order such transitions 
as they occur on the cosmological time scale, such that $T_n^{(a)} > T_n^{(b)}$. The FOPT benchmarks in \cref{tab:properties_GW} achieve the maximal ratio $\Delta v_n/T_n\sim 1$, which is not always correlated with the strength of the PT and the GW peak-amplitude value indicated in the second-to-last column of the table.
In the last column of \cref{tab:properties_GW} we denote by $O(2)$ PTs that are of the second order at tree-level and by $O(1)$ those that are already FOPTs at tree-level. This identification follows our previous work in \cite{Vieu:2018nfq} where we have classified all possible PTs at lowest order in the thermal expansion. We have also took into account the interchange symmetry in the classical field-dependent potential $V_0\left(\phi_\alpha\right)$ given in Eq.~\eqref{eq:V0} to properly identify each transition type. In general $O(1)$-type transitions yield peak amplitudes at or beyond $10^{-16}$ order while $O(2)$-type ones lie at or below this limit. Note that the two highest peaks result from $O(1)$-type PTs producing signals potentially at the reach of LISA. However, while $O(1)$ PTs contribute for larger peak amplitudes they will not necessarily imply observable signatures. For example, there is a third $O(1)$-type transition, $id=6$, (b), which is not strong enough, $\alpha = 0.001$, to generate a peak with an amplitude larger than $h^2 \Omega^\mathrm{peak}_\mathrm{GW} \sim 10^{-16}$. On the other hand, there is a $O(2)$-type transition, $id=2$, (b), which is rather strong, $\alpha = 0.14$, resulting in a comparable peak amplitude of about $h^2 \Omega^\mathrm{peak}_\mathrm{GW} \sim 7\times10^{-16}$.

Consider now the fourth scenario, with $id=4$, in detail (the rightmost line in the right panel of Fig.~\ref{fig:GW_Npeaks}). As in all identified benchmark scenarios, the first transition (a) has a larger frequency than the second one and corresponds to $[0] \to \Phi$, while the second one (b) proceeds with a large $v_s^i \sim v_s^f$ while restoring the EW symmetry (at finite temperature). This may sound counter intuitive since naively one would expect a generation of EW breaking vacua at lower temperature, not EW restoring ones. However, such patterns are indeed possible. The reason is the following: recalling our classification in \cite{Vieu:2018nfq}, point $id=4$ qualifies in a type that we have denoted as HMR-1 if we interchange $\phi_1 \leftrightarrow \phi_s$ in Eq.~\eqref{eq:V0}. Two of the possible transitions are then\footnote{The arrow direction depends on the vacuum energy of each minima at a given temperature.} $(0,0,0) \leftrightarrow (0,0,v_s)$, (a), and $(0,v_2,v_s) \leftrightarrow (0,0,v_s)$, (b), both $O(2)$-type, in agreement with the results found in \cref{tab:properties_GW}.
While the second transition (b) last for about 
ten times longer (compared to the respective Hubble time scale) than the first one (a), the PT strength $\alpha$ is ten times smaller for (a) than (b).

Once again, note that the frequencies of both (a) and (b) FOPTs in the resulting GW spectra for configuration $id=7$ appear rather 
close to each other making the corresponding peaks to partially merge. The resulting broad peak amplitude can be large enough to be probed by proposed space-based interferometers such as BBO and DECIGO. Interestingly enough, we found other examples which correspond to sequential double phase transitions with the following patterns: $[0]\to {\cal H}_2 \to {\cal H}_1$, benchmarks $id=1,3,5,6$, and ${\cal H}_2 \to {\cal H}_{12} \to {\cal H}_1$, benchmark $id=2$, where the second transition (b) is relatively much longer than the first one (a) producing a rather high GW peak at very low frequency. 

It may not be true in general that a peak-amplitude above the sensitivity curve automatically corresponds to an observable signal. This depends on the actual detector configuration, exposure time, source modeling and also on the details of the noise model. To provide a more quantitative information we determine the \textit{Signal to Noise Ratio} (SNR) for two representative scenarios. For instance, we show in \cref{fig:GW_visible} the most pronounced cases which may potentially be at the reach of LISA corresponding to benchmarks $id=3,5$ (green and blue lines, respectively). Focusing on these two cases, and in particular on those corresponding to the higher amplitude peaks, labeled with (b), we use the public online tool \texttt{PTPlot} \cite{Caprini:2019egz} to calculate the SNR for an exposure of 3 and 
7 years showing our results in \cref{tab:SNR}.	
\begin{table}[h!]
	\centering
	\begin{tabular}{|c|c|c|} \hline
		Peak   &  3 year SNR & 7 year SNR 
		\\ \hline
		3 (b)    &  25.1   &  38.4  \\ \hline
		5 (b)    &  10.8   &  16.5  \\ \hline
	\end{tabular}
	\caption{SNR values for the peaks potentially at the reach of LISA.}
	\label{tab:SNR}
\end{table}
Taking an optimistic approach and assuming a minimum $\text{SNR} = 10$ for an observable signal we see that both peaks would indeed be at the reach of LISA readily after three years of data taking. However, if we instead take a conservative approach and follow the criterion $\text{SNR} > 50$ as discussed in \cite{Caprini:2015zlo}, then none of such peaks would be within the LISA range, even after an exposure of 7 years.

From case to case, we observe a large variety of transition patterns and phase 
structures at both (a) and (b) stages. Nevertheless, there are several common features that can be noticed 
for the identified set of double transitions in Table~\ref{tab:properties_GW}. In particular, it is worth mentioning that for all 
the considered benchmark scenarios there are relatively large (but still perturbative) 
scalar self-couplings which enhance thermal scalar masses, also increasing the high-order (particularly, 
$(m/T)^3$) thermal corrections. Indeed, this produces a large enough barrier between the two separate phases (e.g. $[0]$ and $\Phi$ phases 
in the $id=4$ case), turning the second-order tree-level phase transition between them into a first-order one. Recall that from our earlier analysis in \cite{Vieu:2018nfq} the transitions $[0] \to \text{any-phase}$ is of the $O(2)$-type at the lowest order in the thermal expansion. Also, a strong asymmetry between different scalar self-couplings may cause a stronger energy density gradient across the bubble wall, hence causing an effectively stronger transition, thus a larger impact on the primordial GWs. We notice here that at least one counterpart in each (a)+(b) sequence in all considered benchmarks is second-order transition at tree level that becomes a FOPT upon inclusion of relatively 
large higher-order thermal corrections.

For transitions in the same sequence (a)+(b), a smaller frequency typically, although, not exclusively (see $id=7$), corresponds to a larger GW signal, a smaller nucleation temperature $T_n$, often 
a larger $\alpha$ and $v_b$, and a smaller $\beta/H$. For instance, the stronger transitions often correspond to 
a smaller $\beta/H$ in accordance with the full scan data shown in Fig.~\ref{fig:GW_alphabetaH}. This is also related to the fact that in most cases there are two very different types of phase transitions in the same sequence: while the first in the sequence, (a), are typically weak and short lasting becoming first-order via thermal-loop effects, the second ones, (b), are either strong, long lasting and already $O(1)$-type at the leading $(m/T)^2$ order, or strong, short-lasting and $O(2)$-type at the leading order in the thermal expansion. A rich variety of different transition patterns in multi-scalar models 
such as the 2HDSM implies a variety in potential scenarios for sequential phase transitions where correlations and hierarchies between the main characteristics 
are very sensitive to the growing number of model parameters becoming less transparent and predictable.

A natural question is that can we expect more sequential transitions for a given parameter space point of the 2HDSM model under consideration? While we have found a few examples with three sequential transitions all such scenarios have failed the BFB conditions in \cref{BFB}. However, this does not mean that multiple transitions with observable GW spectra can not be found, in particular, for more complicated multi-scalar BSM scenarios where they may become more abundant. It is just getting increasingly harder to identify them technically in such models 
given the growing complexity and dimensionality of the field and parameter spaces. However, it is worth mentioning that in \cite{Addazi:2019dqt} we have observed up to three sequential FOPTs although the simultaneous observations of all three peaks appeared to be rather challenging.  

The observation of multi-peak GW spectra may certainly shed some light on dynamics of the EWPT, particularly, if it is driven by several scalar fields.
The discriminating power for multi-peak GW signatures with respect to the underlining multi-scalar field theory is certainly stronger than for single-peak ones, although harder to experimentally observe. In the considered 2HDSM scenario the GW signals with well-distinguished and potentially detectable peak-amplitudes 
are rather rare. We only found three such configurations in a potentially accessible domain, and all of them with a hardly resolvable 
second peak as illustrated in Fig.~\ref{fig:GW_visible}. A new generation of GW detectors reaching smaller amplitudes and wider frequency domains would be needed for a thorough search for such cosmological events. While further studies are important, the production of two well-separated and potentially detectable (by near-future GW interferometers) peaks in the GW spectrum may be possible in two cases: (i) with an enhanced PT strength due to 
a larger energy budget of EWPTs, and (ii) richer particle spectra typical e.g.~in Grand-unified theories where the loop-induced FOPTs (followed by another very strong FOPT) 
may become strong enough to generate the GW spectra falling within the projected sensitivity limits.

\section{Exotic cosmological events}
\label{Sect:exotic}

Often in the literature, the multi-step transitions are considered to have only one first-order transition step 
which is expected to be much stronger that the other possible steps and thus is typically the only one that 
should be studied (see e.g. Refs.~\citep{Patel:2012pi,Inoue:2015pza}). This is also in accordance with findings 
in the previous section that one of the peaks corresponding to a weaker phase transition has typically 
a much smaller amplitude if the separation between the peak frequencies is large so that the peaks 
are distinguishable. However, under certain requirements on multi-Higgs model parameters, in principle,
there is a possibility to generate strong multi-step transitions already at leading order in thermal expansion
such that several nucleation processes might occur within the same temperature range, e.g.~$\Phi \to {\cal H}_1$ and 
$\Phi \to {\cal H}_2$, yielding the emergence of rather exotic cosmological events. Let us briefly 
consider the possibilities that emerge already in a simple multi-scalar extension of the SM like 
the 2HDSM model discussed above.
\begin{figure}[!h]
\begin{minipage}{0.28\textwidth}
 \centerline{\includegraphics[width=1.2\textwidth]{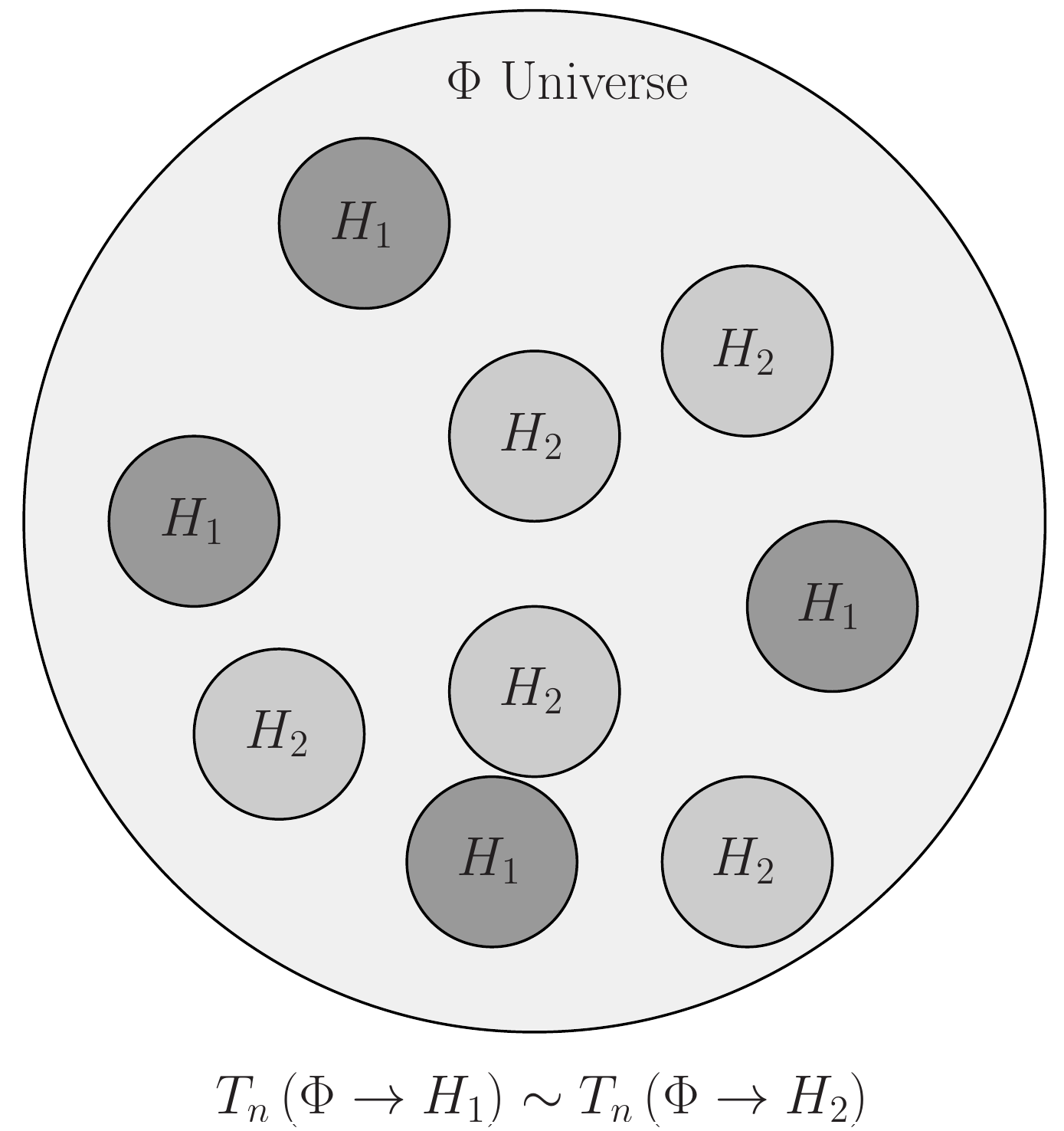}}
\end{minipage}
\hspace{0.6cm}
\begin{minipage}{0.28\textwidth}
 \centerline{\includegraphics[width=1.2\textwidth]{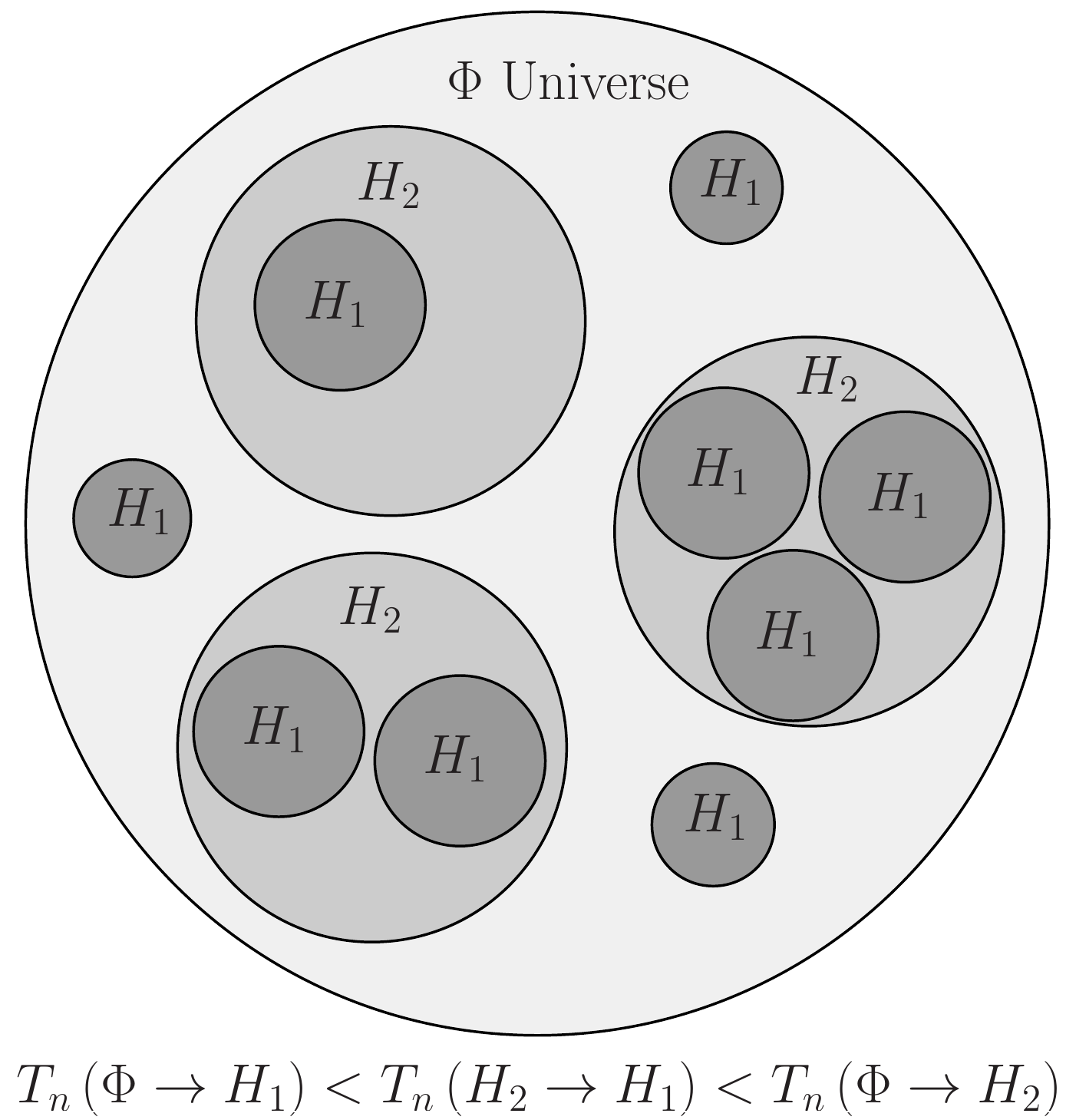}}
\end{minipage}
\hspace{0.6cm}
\begin{minipage}{0.28\textwidth}
 \centerline{\includegraphics[width=1.2\textwidth]{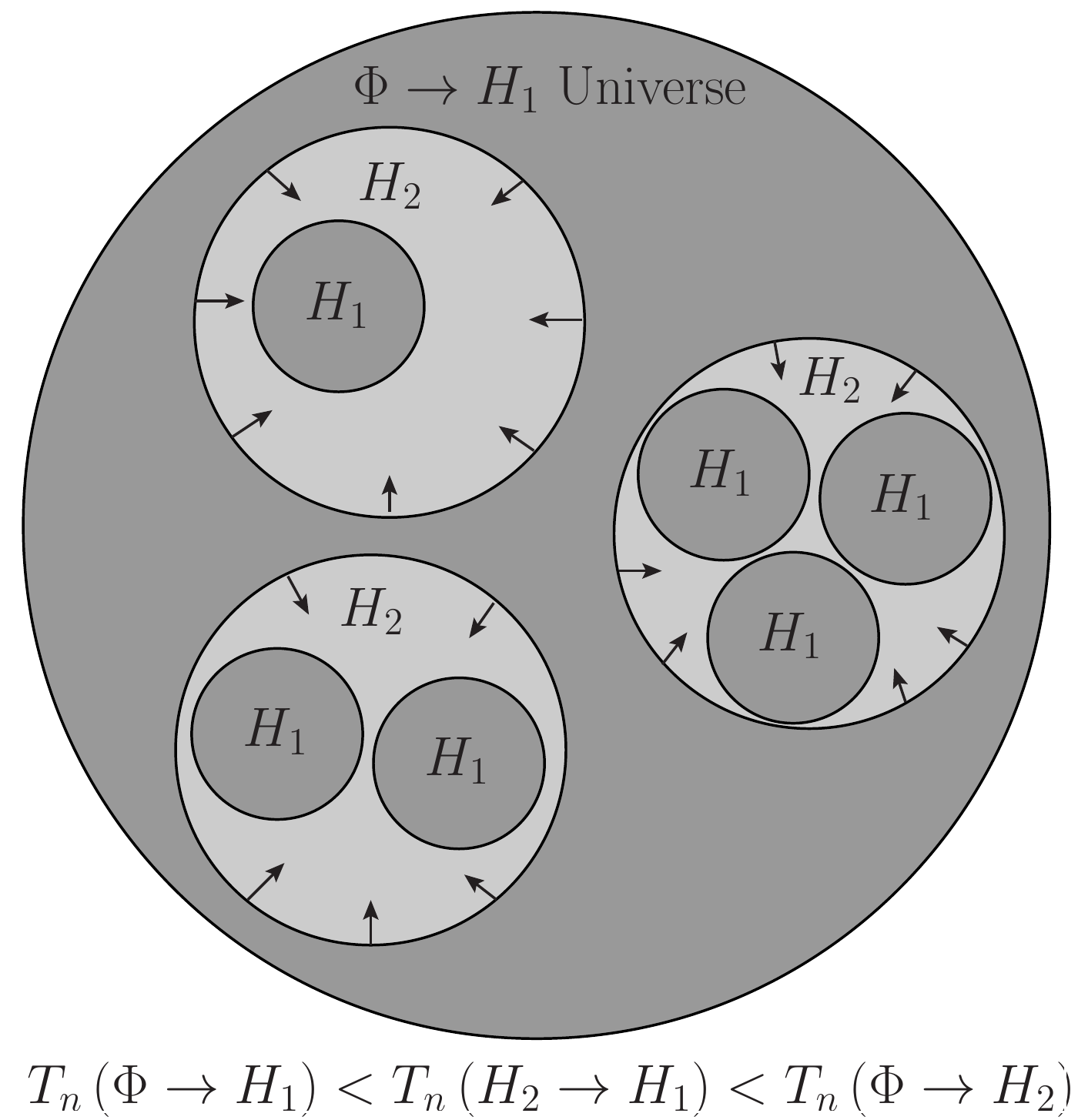}}
\end{minipage}
   \caption{
A schematic illustration of the $\Phi$-phase containing the coexisting ${\cal H}_1$ and ${\cal H}_2$ bubbles
(left panel), and in the nested case of ${\cal H}_1$-bubbles being born inside of ${\cal H}_2$ bubbles (middle panel).
In the right panel, the reoccuring bubbles scenario is shown when ${\cal H}_1$-bubbles nucleation in the $\Phi$ phase
forces the ${\cal H}_2$-bubbles to contract, while ${\cal H}_1$-bubbles are being born inside them.
}
\label{fig:bubbles}
\end{figure}

Quite obviously, different transition sequences could be realized during the same cosmological evolution 
time scale leading to a universe with {\it coexisting bubbles} expanding simultaneously 
(left panel in Fig.~\ref{fig:bubbles}). Indeed, since the effective potential evolves as the temperature of the primordial plasma
drops below $T_n({\cal H}_2 \to {\cal H}_1)$, the initial phase $\Phi$ becomes unstable in the ${\cal H}_1$ direction as well, 
such that the two PTs towards ${\cal H}_1$ and ${\cal H}_2$ phases can occur simultaneously yielding the coexisting bubbles scenario.
At typical temperatures between $T_n(\Phi \to {\cal H}_2) \gtrsim T_n({\cal H}_2 \to {\cal H}_1)$, the ${\cal H}_2$-bubbles nucleate in 
a universe filled with the $\Phi$-phase. Then at $T_n({\cal H}_2 \to {\cal H}_1)$, while the latter are still expanding, 
the ${\cal H}_1$-bubbles emerge and nucleate inside the ${\cal H}_2$-bubbles giving rise to the {\it nested bubbles} 
configuration. As soon as the potential barrier between the phases $\Phi$ and ${\cal H}_1$ vanishes, ${\cal H}_1$-bubbles 
may undergo nucleation in the regions still filled with the $\Phi$-phase. In this case, $\Phi \to {\cal H}_1$ transition washes out 
the $\Phi$-phase outside of the ${\cal H}_2$-bubbles. For an illustration of such a mixed configuration representing the coexistence 
of ${\cal H}_1$ bubbles and nested ${\cal H}_2 \to {\cal H}_1$ phases, see Fig.~\ref{fig:bubbles} (middle panel). Ultimately,
one considers a configuration with the ${\cal H}_1$-bubbles inside the ${\cal H}_2$ ones evolving in a universe containing 
the ${\cal H}_1$-phase, the so-called {\it reoccurring bubble} scenario. Provided that the ${\cal H}_2$-bubbles do not expand into 
the stable ${\cal H}_1$-phase, we expect that they should be pushed inwards and eventually collapse while the ${\cal H}_1$-bubbles 
keep nucleating inside them as shown in Fig.~\ref{fig:bubbles} (right panel). See Ref.~\citep{Vieu:2018zze} for an explicit 
numerical example of such scenarios.

We should of course keep in mind that the nested bubbles could only be nucleated if their nucleation temperatures are 
very close, which makes them unlikely in general. However, this possibility can not be excluded a priori since 
certain symmetries of the high-scale theory may impose specific relations between the model parameters (as it is 
for example the case in Ref.~\citep{Ivanov:2017zjq}) making the exotic objects like the ones discussed above 
theoretically favourable. Since one-step formalism for the primordial GWs spectrum does not apply for sequential transitions 
that have very similar nucleation temperatures, a more sophisticated approach including, in particular, the mutual 
bubble wall collisions, remain to be developed in the future.

\section{Conclusions}
\label{Sect:Conclusions}

We have shown how multi-peaked GW spectra can originate from well-separated multi-step phase transitions 
in multi-Higgs BSM theories. Considering a simple 2HDSM scenario for BSM physics as a suitable benchmark model,
by a detailed numerical scan we have found, classified and described the transition patterns that leads to potentially
observable double-peak configurations. In many identified cases when two subsequent transitions have a different origin, 
i.e.~one is of second order at tree level that becomes a weakly FOPT once higher order corrections are included while the other
is a strong FOPT already at tree level, their combined GW spectrum exhibits two well separated and potentially resolved peaks. 
On the other hand, if sequential phase transitions occur at nearby temperatures, one may expect formation and nucleation of
exotic cosmological objects such as coexisting and nested bubbles. In generic new physics scenarios originating e.g.~from 
Grand-unified field theories, one typically encounters much larger scalar sectors where a more abundant variety of sequential 
phase transition patterns emerge. This leads to potentially observable multi-peaked GW spectra strongly inspiring further 
work in this direction.

\acknowledgments
The authors would like to thank David J.~Weir for insightful clarifications about the SNR calculation in \texttt{PTPlot}.
The authors also thank C.~Herdeiro, M.~Sampaio, J.~Rosa 
and M. Ouerfelli for useful discussions in the various stages of this work.~A.P.M.~is supported by Funda\c{c}\~ao 
para a Ci\^encia e a Tecnologia (FCT), within project UID/MAT/04106/2019 (CIDMA) and by national funds (OE), 
through FCT, I.P., in the scope of the framework contract foreseen in the numbers 4, 5 and 6 of the article 23, 
of the Decree-Law 57/2016, of August 29, changed by Law 57/2017, of July 19.~A.P.M.~is also partially supported by 
the \textit{Enabling Green E-science for the Square Kilometer Array Research Infrastructure} (ENGAGESKA), 
POCI-01-0145-FEDER-022217, and by the project \textit{From Higgs Phenomenology to the Unification 
of Fundamental Interactions}, PTDC/FIS-PAR/31000/2017.~R.P. thanks Prof.~C.~Herdeiro for support of 
the project and hospitality during his visits at Aveiro university.~R.P.~is supported in part by the Swedish 
Research Council grants, contract numbers 621-2013-4287 and 2016-05996, by the Ministry of Education, 
Youth and  Sports of the Czech Republic, project LT17018, as well as by the European Research Council (ERC) 
under the European Union's Horizon 2020 research and innovation programme (grant agreement No 668679).

\appendix

\section{Efficiency coefficients}
\label{sec:App-A}

In this appendix we would like to provide semi-analytical expressions for the efficiency coefficients relevant for our studies. Recalling that we are interested in detonations, \cref{eq:Opeak1,eq:Opeak2,eq:spectrum} are valid for bubble wall velocities above the Chapman-Jouguet speed, $v_\mathrm{J}$, where the fraction of vacuum energy that is converted into kinetic energy reads
\begin{equation}
	\kappa = \dfrac{(v_\mathrm{J}-1)^3 v_\mathrm{J}^{5/2} v_\mathrm{b}^{-5/2} \kappa_1 \kappa_2 }{\left[(v_\mathrm{J}-1)^3 - (v_\mathrm{b}-1)^3\right] v_\mathrm{J}^{5/2} \kappa_1 + (v_\mathrm{b}-1)^3 \kappa_2}
	\label{eq:kappa}
\end{equation} 
with $\kappa_1$ the efficiency factor for the limit of Jouguet detonations, i.e.~$v_\mathrm{b} = v_\mathrm{J}$
\begin{equation}
	\kappa_1 = \dfrac{\sqrt{\alpha}}{0.135+\sqrt{0.98 + \alpha}}\,,
\end{equation}
$\kappa_2$ the efficiency factor for very large bubble wall velocities, i.e.~$v_\mathrm{b} \to 1$
\begin{equation}
	\kappa_2 = \dfrac{\alpha}{0.73 + 0.083 \sqrt{\alpha}+ \alpha}
\end{equation}
and
$v_\mathrm{J}$ the Chapman-Jouguet speed
\begin{equation}
	v_\mathrm{J} = \dfrac{1}{1+\alpha} \left(c_s + \sqrt{\alpha^2 + \tfrac{2}{3} \alpha}\right)\,.
	\label{eq:vJ}
\end{equation}


\bibliographystyle{JHEP}
\bibliography{bib}

\providecommand{\href}[2]{#2}\begingroup\raggedright\begin{thebibliography}{100}

\bibitem{Aad:2012tfa}
{\bf ATLAS} Collaboration, G.~Aad {\em et~al.}, {\it {Observation of a new
  particle in the search for the Standard Model Higgs boson with the ATLAS
  detector at the LHC}},  {\em Phys. Lett.} {\bf B716} (2012) 1--29,
  [\href{https://arxiv.org/abs/1207.7214}{{\tt 1207.7214}}].

\bibitem{Chatrchyan:2012xdj}
{\bf CMS} Collaboration, S.~Chatrchyan {\em et~al.}, {\it {Observation of a new
  boson at a mass of 125 GeV with the CMS experiment at the LHC}},  {\em Phys.
  Lett.} {\bf B716} (2012) 30--61, [\href{https://arxiv.org/abs/1207.7235}{{\tt
  1207.7235}}].

\bibitem{Maggiore:2018sht}
M.~Maggiore, {\em {Gravitational Waves. Vol. 2}}.
\newblock Oxford University Press, 2018.

\bibitem{Caprini:2018mtu}
C.~Caprini and D.~G. Figueroa, {\it {Cosmological Backgrounds of Gravitational
  Waves}},  {\em Class. Quant. Grav.} {\bf 35} (2018), no.~16 163001,
  [\href{https://arxiv.org/abs/1801.04268}{{\tt 1801.04268}}].

\bibitem{Mazumdar:2018dfl}
A.~Mazumdar and G.~White, {\it {Review of cosmic phase transitions: their
  significance and experimental signatures}},  {\em Rept. Prog. Phys.} {\bf 82}
  (2019), no.~7 076901, [\href{https://arxiv.org/abs/1811.01948}{{\tt
  1811.01948}}].

\bibitem{Hashino:2018wee}
K.~Hashino, R.~Jinno, M.~Kakizaki, S.~Kanemura, T.~Takahashi, and M.~Takimoto,
  {\it {Selecting models of first-order phase transitions using the synergy
  between collider and gravitational-wave experiments}},  {\em Phys. Rev.} {\bf
  D99} (2019), no.~7 075011, [\href{https://arxiv.org/abs/1809.04994}{{\tt
  1809.04994}}].

\bibitem{Audley:2017drz}
{\bf LISA} Collaboration, P.~Amaro-Seoane {\em et~al.}, {\it {Laser
  Interferometer Space Antenna}},  \href{https://arxiv.org/abs/1702.00786}{{\tt
  1702.00786}}.

\bibitem{Seto:2001qf}
N.~Seto, S.~Kawamura, and T.~Nakamura, {\it {Possibility of direct measurement
  of the acceleration of the universe using 0.1-Hz band laser interferometer
  gravitational wave antenna in space}},  {\em Phys. Rev. Lett.} {\bf 87}
  (2001) 221103, [\href{https://arxiv.org/abs/astro-ph/0108011}{{\tt
  astro-ph/0108011}}].

\bibitem{Kudoh:2005as}
H.~Kudoh, A.~Taruya, T.~Hiramatsu, and Y.~Himemoto, {\it {Detecting a
  gravitational-wave background with next-generation space interferometers}},
  {\em Phys. Rev.} {\bf D73} (2006) 064006,
  [\href{https://arxiv.org/abs/gr-qc/0511145}{{\tt gr-qc/0511145}}].

\bibitem{Kawamura:2011zz}
S.~Kawamura {\em et~al.}, {\it {The Japanese space gravitational wave antenna:
  DECIGO}},  {\em Class. Quant. Grav.} {\bf 28} (2011) 094011.

\bibitem{Kuroyanagi:2014qaa}
S.~Kuroyanagi, S.~Tsujikawa, T.~Chiba, and N.~Sugiyama, {\it {Implications of
  the B-mode Polarization Measurement for Direct Detection of Inflationary
  Gravitational Waves}},  {\em Phys. Rev.} {\bf D90} (2014), no.~6 063513,
  [\href{https://arxiv.org/abs/1406.1369}{{\tt 1406.1369}}].

\bibitem{Crowder:2005nr}
J.~Crowder and N.~J. Cornish, {\it {Beyond LISA: Exploring future gravitational
  wave missions}},  {\em Phys. Rev.} {\bf D72} (2005) 083005,
  [\href{https://arxiv.org/abs/gr-qc/0506015}{{\tt gr-qc/0506015}}].

\bibitem{Corbin:2005ny}
V.~Corbin and N.~J. Cornish, {\it {Detecting the cosmic gravitational wave
  background with the big bang observer}},  {\em Class. Quant. Grav.} {\bf 23}
  (2006) 2435--2446, [\href{https://arxiv.org/abs/gr-qc/0512039}{{\tt
  gr-qc/0512039}}].

\bibitem{Huang:2016cjm}
P.~Huang, A.~J. Long, and L.-T. Wang, {\it {Probing the Electroweak Phase
  Transition with Higgs Factories and Gravitational Waves}},  {\em Phys. Rev.}
  {\bf D94} (2016), no.~7 075008, [\href{https://arxiv.org/abs/1608.06619}{{\tt
  1608.06619}}].

\bibitem{No:2011fi}
J.~M. No, {\it {Large Gravitational Wave Background Signals in Electroweak
  Baryogenesis Scenarios}},  {\em Phys. Rev.} {\bf D84} (2011) 124025,
  [\href{https://arxiv.org/abs/1103.2159}{{\tt 1103.2159}}].

\bibitem{Grojean:2006bp}
C.~Grojean and G.~Servant, {\it {Gravitational Waves from Phase Transitions at
  the Electroweak Scale and Beyond}},  {\em Phys. Rev.} {\bf D75} (2007)
  043507, [\href{https://arxiv.org/abs/hep-ph/0607107}{{\tt hep-ph/0607107}}].

\bibitem{Apreda:2001us}
R.~Apreda, M.~Maggiore, A.~Nicolis, and A.~Riotto, {\it {Gravitational waves
  from electroweak phase transitions}},  {\em Nucl. Phys.} {\bf B631} (2002)
  342--368, [\href{https://arxiv.org/abs/gr-qc/0107033}{{\tt gr-qc/0107033}}].

\bibitem{Hashino:2016rvx}
K.~Hashino, M.~Kakizaki, S.~Kanemura, and T.~Matsui, {\it {Synergy between
  measurements of gravitational waves and the triple-Higgs coupling in probing
  the first-order electroweak phase transition}},  {\em Phys. Rev.} {\bf D94}
  (2016), no.~1 015005, [\href{https://arxiv.org/abs/1604.02069}{{\tt
  1604.02069}}].

\bibitem{Hashino:2016xoj}
K.~Hashino, M.~Kakizaki, S.~Kanemura, P.~Ko, and T.~Matsui, {\it {Gravitational
  waves and Higgs boson couplings for exploring first order phase transition in
  the model with a singlet scalar field}},  {\em Phys. Lett.} {\bf B766} (2017)
  49--54, [\href{https://arxiv.org/abs/1609.00297}{{\tt 1609.00297}}].

\bibitem{Kakizaki:2015wua}
M.~Kakizaki, S.~Kanemura, and T.~Matsui, {\it {Gravitational waves as a probe
  of extended scalar sectors with the first order electroweak phase
  transition}},  {\em Phys. Rev.} {\bf D92} (2015), no.~11 115007,
  [\href{https://arxiv.org/abs/1509.08394}{{\tt 1509.08394}}].

\bibitem{Dev:2016feu}
P.~S.~B. Dev and A.~Mazumdar, {\it {Probing the Scale of New Physics by
  Advanced LIGO/VIRGO}},  {\em Phys. Rev.} {\bf D93} (2016), no.~10 104001,
  [\href{https://arxiv.org/abs/1602.04203}{{\tt 1602.04203}}].

\bibitem{Dev:2016hxv}
P.~S.~B. Dev, M.~Lindner, and S.~Ohmer, {\it {Gravitational waves as a new
  probe of Bose–Einstein condensate Dark Matter}},  {\em Phys. Lett.} {\bf
  B773} (2017) 219--224, [\href{https://arxiv.org/abs/1609.03939}{{\tt
  1609.03939}}].

\bibitem{Addazi:2018nzm}
A.~Addazi, A.~Marcianò, and R.~Pasechnik, {\it {Probing Trans-electroweak
  First Order Phase Transitions from Gravitational Waves}},  {\em MDPI Physics}
  {\bf 1} (2019), no.~1 92--102, [\href{https://arxiv.org/abs/1811.09074}{{\tt
  1811.09074}}].

\bibitem{Vieu:2018zze}
T.~Vieu, A.~P. Morais, and R.~Pasechnik, {\it {Multi-peaked signatures of
  primordial gravitational waves from multi-step electroweak phase
  transition}},  \href{https://arxiv.org/abs/1802.10109}{{\tt 1802.10109}}.

\bibitem{Angelescu:2018dkk}
A.~Angelescu and P.~Huang, {\it {Multistep Strongly First Order Phase
  Transitions from New Fermions at the TeV Scale}},  {\em Phys. Rev.} {\bf D99}
  (2019), no.~5 055023, [\href{https://arxiv.org/abs/1812.08293}{{\tt
  1812.08293}}].

\bibitem{Alanne:2019bsm}
T.~Alanne, T.~Hugle, M.~Platscher, and K.~Schmitz, {\it {A fresh look at the
  gravitational-wave signal from cosmological phase transitions}},
  \href{https://arxiv.org/abs/1909.11356}{{\tt 1909.11356}}.

\bibitem{Addazi:2019dqt}
A.~Addazi, A.~Marcianò, A.~P. Morais, R.~Pasechnik, R.~Srivastava, and
  J.~W.~F. Valle, {\it {Gravitational footprints of massive neutrinos and
  lepton number breaking}},  \href{https://arxiv.org/abs/1909.09740}{{\tt
  1909.09740}}.

\bibitem{Mohamadnejad:2019vzg}
A.~Mohamadnejad, {\it {Gravitational waves from scale-invariant vector dark
  matter model: Probing below the neutrino-floor}},
  \href{https://arxiv.org/abs/1907.08899}{{\tt 1907.08899}}.

\bibitem{Alves:2018jsw}
A.~Alves, T.~Ghosh, H.-K. Guo, K.~Sinha, and D.~Vagie, {\it {Collider and
  Gravitational Wave Complementarity in Exploring the Singlet Extension of the
  Standard Model}},  {\em JHEP} {\bf 04} (2019) 052,
  [\href{https://arxiv.org/abs/1812.09333}{{\tt 1812.09333}}].

\bibitem{Alves:2018oct}
A.~Alves, T.~Ghosh, H.-K. Guo, and K.~Sinha, {\it {Resonant Di-Higgs Production
  at Gravitational Wave Benchmarks: A Collider Study using Machine Learning}},
  {\em JHEP} {\bf 12} (2018) 070, [\href{https://arxiv.org/abs/1808.08974}{{\tt
  1808.08974}}].

\bibitem{Chao:2017ilw}
W.~Chao, W.-F. Cui, H.-K. Guo, and J.~Shu, {\it {Gravitational Wave Imprint of
  New Symmetry Breaking}},  \href{https://arxiv.org/abs/1707.09759}{{\tt
  1707.09759}}.

\bibitem{Bian:2019szo}
L.~Bian, W.~Cheng, H.-K. Guo, and Y.~Zhang, {\it {Gravitational waves triggered
  by $B-L$ charged hidden scalar and leptogenesis}},
  \href{https://arxiv.org/abs/1907.13589}{{\tt 1907.13589}}.

\bibitem{Dev:2019njv}
P.~S.~B. Dev, F.~Ferrer, Y.~Zhang, and Y.~Zhang, {\it {Gravitational Waves from
  First-Order Phase Transition in a Simple Axion-Like Particle Model}},
  \href{https://arxiv.org/abs/1905.00891}{{\tt 1905.00891}}.

\bibitem{Wang:2019pet}
X.~Wang, F.~P. Huang, and X.~Zhang, {\it {Gravitational wave and collider
  signals in complex two-Higgs doublet model with dynamical CP-violation at
  finite temperature}},  \href{https://arxiv.org/abs/1909.02978}{{\tt
  1909.02978}}.

\bibitem{Kosowsky:1991ua}
A.~Kosowsky, M.~S. Turner, and R.~Watkins, {\it {Gravitational radiation from
  colliding vacuum bubbles}},  {\em Phys. Rev.} {\bf D45} (1992) 4514--4535.

\bibitem{Kosowsky:1992rz}
A.~Kosowsky, M.~S. Turner, and R.~Watkins, {\it {Gravitational waves from first
  order cosmological phase transitions}},  {\em Phys. Rev. Lett.} {\bf 69}
  (1992) 2026--2029.

\bibitem{Hindmarsh:2013xza}
M.~Hindmarsh, S.~J. Huber, K.~Rummukainen, and D.~J. Weir, {\it {Gravitational
  waves from the sound of a first order phase transition}},  {\em Phys. Rev.
  Lett.} {\bf 112} (2014) 041301, [\href{https://arxiv.org/abs/1304.2433}{{\tt
  1304.2433}}].

\bibitem{Hindmarsh:2015qta}
M.~Hindmarsh, S.~J. Huber, K.~Rummukainen, and D.~J. Weir, {\it {Numerical
  simulations of acoustically generated gravitational waves at a first order
  phase transition}},  {\em Phys. Rev.} {\bf D92} (2015), no.~12 123009,
  [\href{https://arxiv.org/abs/1504.03291}{{\tt 1504.03291}}].

\bibitem{Sakharov:1967dj}
A.~D. Sakharov, {\it {Violation of CP Invariance, c Asymmetry, and Baryon
  Asymmetry of the Universe}},  {\em Pisma Zh. Eksp. Teor. Fiz.} {\bf 5} (1967)
  32--35. [Usp. Fiz. Nauk161,61(1991)].

\bibitem{Branco:2011iw}
G.~C. Branco, P.~M. Ferreira, L.~Lavoura, M.~N. Rebelo, M.~Sher, and J.~P.
  Silva, {\it {Theory and phenomenology of two-Higgs-doublet models}},  {\em
  Phys. Rept.} {\bf 516} (2012) 1--102,
  [\href{https://arxiv.org/abs/1106.0034}{{\tt 1106.0034}}].

\bibitem{Barger:2007im}
V.~Barger, P.~Langacker, M.~McCaskey, M.~J. Ramsey-Musolf, and G.~Shaughnessy,
  {\it {LHC Phenomenology of an Extended Standard Model with a Real Scalar
  Singlet}},  {\em Phys. Rev.} {\bf D77} (2008) 035005,
  [\href{https://arxiv.org/abs/0706.4311}{{\tt 0706.4311}}].

\bibitem{Barger:2008jx}
V.~Barger, P.~Langacker, M.~McCaskey, M.~Ramsey-Musolf, and G.~Shaughnessy,
  {\it {Complex Singlet Extension of the Standard Model}},  {\em Phys. Rev.}
  {\bf D79} (2009) 015018, [\href{https://arxiv.org/abs/0811.0393}{{\tt
  0811.0393}}].

\bibitem{Chala:2016ykx}
M.~Chala, G.~Nardini, and I.~Sobolev, {\it {Unified explanation for dark matter
  and electroweak baryogenesis with direct detection and gravitational wave
  signatures}},  {\em Phys. Rev.} {\bf D94} (2016), no.~5 055006,
  [\href{https://arxiv.org/abs/1605.08663}{{\tt 1605.08663}}].

\bibitem{Vaskonen:2016yiu}
V.~Vaskonen, {\it {Electroweak baryogenesis and gravitational waves from a real
  scalar singlet}},  {\em Phys. Rev.} {\bf D95} (2017), no.~12 123515,
  [\href{https://arxiv.org/abs/1611.02073}{{\tt 1611.02073}}].

\bibitem{Beniwal:2017eik}
A.~Beniwal, M.~Lewicki, J.~D. Wells, M.~White, and A.~G. Williams, {\it
  {Gravitational wave, collider and dark matter signals from a scalar singlet
  electroweak baryogenesis}},  \href{https://arxiv.org/abs/1702.06124}{{\tt
  1702.06124}}.

\bibitem{Cline:2012hg}
J.~M. Cline and K.~Kainulainen, {\it {Electroweak baryogenesis and dark matter
  from a singlet Higgs}},  {\em JCAP} {\bf 1301} (2013) 012,
  [\href{https://arxiv.org/abs/1210.4196}{{\tt 1210.4196}}].

\bibitem{Kurup:2017dzf}
G.~Kurup and M.~Perelstein, {\it {Dynamics of Electroweak Phase Transition In
  Singlet-Scalar Extension of the Standard Model}},  {\em Phys. Rev.} {\bf D96}
  (2017) 015036, [\href{https://arxiv.org/abs/1704.03381}{{\tt 1704.03381}}].

\bibitem{Li:2014wia}
T.~Li and Y.-F. Zhou, {\it {Strongly first order phase transition in the
  singlet fermionic dark matter model after LUX}},  {\em JHEP} {\bf 07} (2014)
  006, [\href{https://arxiv.org/abs/1402.3087}{{\tt 1402.3087}}].

\bibitem{Jiang:2015cwa}
M.~Jiang, L.~Bian, W.~Huang, and J.~Shu, {\it {Impact of a complex singlet:
  Electroweak baryogenesis and dark matter}},  {\em Phys. Rev.} {\bf D93}
  (2016), no.~6 065032, [\href{https://arxiv.org/abs/1502.07574}{{\tt
  1502.07574}}].

\bibitem{Basler:2016obg}
P.~Basler, M.~Krause, M.~Muhlleitner, J.~Wittbrodt, and A.~Wlotzka, {\it
  {Strong First Order Electroweak Phase Transition in the CP-Conserving 2HDM
  Revisited}},  {\em JHEP} {\bf 02} (2017) 121,
  [\href{https://arxiv.org/abs/1612.04086}{{\tt 1612.04086}}].

\bibitem{Basler:2017uxn}
P.~Basler, M.~Mühlleitner, and J.~Wittbrodt, {\it {The CP-Violating 2HDM in
  Light of a Strong First Order Electroweak Phase Transition and Implications
  for Higgs Pair Production}},  \href{https://arxiv.org/abs/1711.04097}{{\tt
  1711.04097}}.

\bibitem{Dorsch:2013wja}
G.~C. Dorsch, S.~J. Huber, and J.~M. No, {\it {A strong electroweak phase
  transition in the 2HDM after LHC8}},  {\em JHEP} {\bf 10} (2013) 029,
  [\href{https://arxiv.org/abs/1305.6610}{{\tt 1305.6610}}].

\bibitem{Ginzburg:2010wa}
I.~F. Ginzburg, K.~A. Kanishev, M.~Krawczyk, and D.~Sokolowska, {\it {Evolution
  of Universe to the present inert phase}},  {\em Phys. Rev.} {\bf D82} (2010)
  123533, [\href{https://arxiv.org/abs/1009.4593}{{\tt 1009.4593}}].

\bibitem{Chala:2018opy}
M.~Chala, M.~Ramos, and M.~Spannowsky, {\it {Gravitational wave and collider
  probes of a triplet Higgs sector with a low cutoff}},  {\em Eur. Phys. J.}
  {\bf C79} (2019), no.~2 156, [\href{https://arxiv.org/abs/1812.01901}{{\tt
  1812.01901}}].

\bibitem{Bian:2017wfv}
L.~Bian, H.-K. Guo, and J.~Shu, {\it {Gravitational Waves, baryon asymmetry of
  the universe and electric dipole moment in the CP-violating NMSSM}},  {\em
  Chin. Phys.} {\bf C42} (2018), no.~9 093106,
  [\href{https://arxiv.org/abs/1704.02488}{{\tt 1704.02488}}].

\bibitem{Chao:2017vrq}
W.~Chao, H.-K. Guo, and J.~Shu, {\it {Gravitational Wave Signals of Electroweak
  Phase Transition Triggered by Dark Matter}},  {\em JCAP} {\bf 1709} (2017),
  no.~09 009, [\href{https://arxiv.org/abs/1702.02698}{{\tt 1702.02698}}].

\bibitem{Patel:2012pi}
H.~H. Patel and M.~J. Ramsey-Musolf, {\it {Stepping Into Electroweak Symmetry
  Breaking: Phase Transitions and Higgs Phenomenology}},  {\em Phys. Rev.} {\bf
  D88} (2013) 035013, [\href{https://arxiv.org/abs/1212.5652}{{\tt
  1212.5652}}].

\bibitem{Inoue:2015pza}
S.~Inoue, G.~Ovanesyan, and M.~J. Ramsey-Musolf, {\it {Two-Step Electroweak
  Baryogenesis}},  {\em Phys. Rev.} {\bf D93} (2016) 015013,
  [\href{https://arxiv.org/abs/1508.05404}{{\tt 1508.05404}}].

\bibitem{Blinov:2015sna}
N.~Blinov, J.~Kozaczuk, D.~E. Morrissey, and C.~Tamarit, {\it {Electroweak
  Baryogenesis from Exotic Electroweak Symmetry Breaking}},  {\em Phys. Rev.}
  {\bf D92} (2015), no.~3 035012, [\href{https://arxiv.org/abs/1504.05195}{{\tt
  1504.05195}}].

\bibitem{Ramsey-Musolf:2017tgh}
M.~J. Ramsey-Musolf, G.~White, and P.~Winslow, {\it {Color Breaking
  Baryogenesis}},  \href{https://arxiv.org/abs/1708.07511}{{\tt 1708.07511}}.

\bibitem{Huang:2017laj}
F.~P. Huang and X.~Zhang, {\it {Probing the gauge symmetry breaking of the
  early universe in 3-3-1 models and beyond by gravitational waves}},  {\em
  Phys. Lett.} {\bf B788} (2019) 288--294,
  [\href{https://arxiv.org/abs/1701.04338}{{\tt 1701.04338}}].

\bibitem{Ashoorioon:2009nf}
A.~Ashoorioon and T.~Konstandin, {\it {Strong electroweak phase transitions
  without collider traces}},  {\em JHEP} {\bf 07} (2009) 086,
  [\href{https://arxiv.org/abs/0904.0353}{{\tt 0904.0353}}].

\bibitem{Alanne:2016wtx}
T.~Alanne, K.~Kainulainen, K.~Tuominen, and V.~Vaskonen, {\it {Baryogenesis in
  the two doublet and inert singlet extension of the Standard Model}},  {\em
  JCAP} {\bf 1608} (2016), no.~08 057,
  [\href{https://arxiv.org/abs/1607.03303}{{\tt 1607.03303}}].

\bibitem{Kang:2017mkl}
Z.~Kang, P.~Ko, and T.~Matsui, {\it {Strong first order EWPT \& strong
  gravitational waves in Z$_{3}$-symmetric singlet scalar extension}},  {\em
  JHEP} {\bf 02} (2018) 115, [\href{https://arxiv.org/abs/1706.09721}{{\tt
  1706.09721}}].

\bibitem{Vieu:2018nfq}
T.~Vieu, A.~P. Morais, and R.~Pasechnik, {\it {Electroweak phase transitions in
  multi-Higgs models: the case of Trinification-inspired THDSM}},  {\em JCAP}
  {\bf 1807} (2018), no.~07 014, [\href{https://arxiv.org/abs/1801.02670}{{\tt
  1801.02670}}].

\bibitem{Chung:2010cd}
D.~J.~H. Chung and A.~J. Long, {\it {Electroweak Phase Transition in the
  munuSSM}},  {\em Phys. Rev.} {\bf D81} (2010) 123531,
  [\href{https://arxiv.org/abs/1004.0942}{{\tt 1004.0942}}].

\bibitem{Camargo-Molina:2016yqm}
J.~E. Camargo-Molina, A.~P. Morais, A.~Ordell, R.~Pasechnik, M.~O. Sampaio, and
  J.~Wessén, {\it {Reviving trinification models through an E6 -extended
  supersymmetric GUT}},  {\em Phys. Rev.} {\bf D95} (2017), no.~7 075031,
  [\href{https://arxiv.org/abs/1610.03642}{{\tt 1610.03642}}].

\bibitem{Camargo-Molina:2017kxd}
J.~E. Camargo-Molina, A.~P. Morais, A.~Ordell, R.~Pasechnik, and J.~Wessén,
  {\it {Scale hierarchies, symmetry breaking and SM-like fermions in
  $\mathrm{SU}(3)$-family extended SUSY trinification}},
  \href{https://arxiv.org/abs/1711.05199}{{\tt 1711.05199}}.

\bibitem{Morais:2020odg}
A.~P. Morais, R.~Pasechnik, and W.~Porod, {\it {Grand Unified origin of gauge
  interactions and families replication in the Standard Model}},
  \href{https://arxiv.org/abs/2001.04804}{{\tt 2001.04804}}.

\bibitem{Morais:2020ypd}
A.~P. Morais, R.~Pasechnik, and W.~Porod, {\it {Prospects for New Physics from
  gauge Left-Right-Colour-Family Grand Unification}},
  \href{https://arxiv.org/abs/2001.06383}{{\tt 2001.06383}}.

\bibitem{Berezinsky:1993fm}
V.~Berezinsky and J.~W.~F. Valle, {\it {The KeV majoron as a dark matter
  particle}},  {\em Phys. Lett.} {\bf B318} (1993) 360--366,
  [\href{https://arxiv.org/abs/hep-ph/9309214}{{\tt hep-ph/9309214}}].

\bibitem{Lattanzi:2007ux}
M.~Lattanzi and J.~W.~F. Valle, {\it {Decaying warm dark matter and neutrino
  masses}},  {\em Phys. Rev. Lett.} {\bf 99} (2007) 121301,
  [\href{https://arxiv.org/abs/0705.2406}{{\tt 0705.2406}}].

\bibitem{Kuo:2018fgw}
J.-L. Kuo {\em et~al.}, {\it {Decaying warm dark matter and structure
  formation}},  {\em JCAP} {\bf 1812} (2018), no.~12 026,
  [\href{https://arxiv.org/abs/1803.05650}{{\tt 1803.05650}}].

\bibitem{Lattanzi:2013uza}
M.~Lattanzi, S.~Riemer-Sorensen, M.~Tortola, and J.~W.~F. Valle, {\it {Updated
  CMB and x- and $\gamma$-ray constraints on Majoron dark matter}},  {\em Phys.
  Rev.} {\bf D88} (2013), no.~6 063528,
  [\href{https://arxiv.org/abs/1303.4685}{{\tt 1303.4685}}].

\bibitem{Bazzocchi:2008fh}
F.~Bazzocchi {\em et~al.}, {\it {X-ray photons from late-decaying majoron dark
  matter}},  {\em JCAP} {\bf 0808} (2008) 013,
  [\href{https://arxiv.org/abs/0805.2372}{{\tt 0805.2372}}].

\bibitem{Kannike:2012pe}
K.~Kannike, {\it {Vacuum Stability Conditions From Copositivity Criteria}},
  {\em Eur. Phys. J.} {\bf C72} (2012) 2093,
  [\href{https://arxiv.org/abs/1205.3781}{{\tt 1205.3781}}].

\bibitem{Quiros:1999jp}
M.~Quiros, {\it {Finite temperature field theory and phase transitions}},  {\em
  Proceedings of Summer School in High-Energy Physics and Cosmology: Trieste,
  Italy, June 29-July 17, 1998} (1999) 187--259,
  [\href{https://arxiv.org/abs/hep-ph/9901312}{{\tt hep-ph/9901312}}].

\bibitem{Curtin:2016urg}
D.~Curtin, P.~Meade, and H.~Ramani, {\it {Thermal Resummation and Phase
  Transitions}},  \href{https://arxiv.org/abs/1612.00466}{{\tt 1612.00466}}.

\bibitem{Dolan:1973qd}
L.~Dolan and R.~Jackiw, {\it {Symmetry Behavior at Finite Temperature}},  {\em
  Phys. Rev.} {\bf D9} (1974) 3320--3341.

\bibitem{Parwani:1991gq}
R.~R. Parwani, {\it {Resummation in a hot scalar field theory}},  {\em Phys.
  Rev.} {\bf D45} (1992) 4695,
  [\href{https://arxiv.org/abs/hep-ph/9204216}{{\tt hep-ph/9204216}}].
  [Erratum: Phys. Rev.D48,5965(1993)].

\bibitem{Arnold:1992rz}
P.~B. Arnold and O.~Espinosa, {\it {The Effective potential and first order
  phase transitions: Beyond leading-order}},  {\em Phys. Rev.} {\bf D47} (1993)
  3546, [\href{https://arxiv.org/abs/hep-ph/9212235}{{\tt hep-ph/9212235}}].
  [Erratum: Phys. Rev.D50,6662(1994)].

\bibitem{Espinosa:1995se}
J.~R. Espinosa and M.~Quiros, {\it {Improved metastability bounds on the
  standard model Higgs mass}},  {\em Phys. Lett.} {\bf B353} (1995) 257--266,
  [\href{https://arxiv.org/abs/hep-ph/9504241}{{\tt hep-ph/9504241}}].

\bibitem{Linde1983}
A.~Linde, {\it Decay of the false vacuum at finite temperature},  {\em Nuclear
  Physics B} {\bf 216} (1983), no.~2 421 -- 445.

\bibitem{Dine:1992wr}
M.~Dine, R.~G. Leigh, P.~Y. Huet, A.~D. Linde, and D.~A. Linde, {\it {Towards
  the theory of the electroweak phase transition}},  {\em Phys. Rev.} {\bf D46}
  (1992) 550--571, [\href{https://arxiv.org/abs/hep-ph/9203203}{{\tt
  hep-ph/9203203}}].

\bibitem{Coleman:1977py}
S.~R. Coleman, {\it {The Fate of the False Vacuum. 1. Semiclassical Theory}},
  {\em Phys. Rev.} {\bf D15} (1977) 2929--2936. [Erratum: Phys.
  Rev.D16,1248(1977)].

\bibitem{Wainwright:2011kj}
C.~L. Wainwright, {\it {CosmoTransitions: Computing Cosmological Phase
  Transition Temperatures and Bubble Profiles with Multiple Fields}},  {\em
  Comput. Phys. Commun.} {\bf 183} (2012) 2006--2013,
  [\href{https://arxiv.org/abs/1109.4189}{{\tt 1109.4189}}].

\bibitem{Kuzmin:1985mm}
V.~A. Kuzmin, V.~A. Rubakov, and M.~E. Shaposhnikov, {\it {On the Anomalous
  Electroweak Baryon Number Nonconservation in the Early Universe}},  {\em
  Phys. Lett.} {\bf 155B} (1985) 36.

\bibitem{Ahriche:2014jna}
A.~Ahriche, T.~A. Chowdhury, and S.~Nasri, {\it {Sphalerons and the Electroweak
  Phase Transition in Models with Higher Scalar Representations}},  {\em JHEP}
  {\bf 11} (2014) 096, [\href{https://arxiv.org/abs/1409.4086}{{\tt
  1409.4086}}].

\bibitem{Patel:2011th}
H.~H. Patel and M.~J. Ramsey-Musolf, {\it {Baryon Washout, Electroweak Phase
  Transition, and Perturbation Theory}},  {\em JHEP} {\bf 07} (2011) 029,
  [\href{https://arxiv.org/abs/1101.4665}{{\tt 1101.4665}}].

\bibitem{Nielsen:1975fs}
N.~K. Nielsen, {\it {On the Gauge Dependence of Spontaneous Symmetry Breaking
  in Gauge Theories}},  {\em Nucl. Phys.} {\bf B101} (1975) 173--188.

\bibitem{Chiang:2017zbz}
C.-W. Chiang and E.~Senaha, {\it {On gauge dependence of gravitational waves
  from a first-order phase transition in classical scale-invariant $U(1)'$
  models}},  {\em Phys. Lett.} {\bf B774} (2017) 489--493,
  [\href{https://arxiv.org/abs/1707.06765}{{\tt 1707.06765}}].

\bibitem{Wainwright:2011qy}
C.~Wainwright, S.~Profumo, and M.~J. Ramsey-Musolf, {\it {Gravity Waves from a
  Cosmological Phase Transition: Gauge Artifacts and Daisy Resummations}},
  {\em Phys. Rev.} {\bf D84} (2011) 023521,
  [\href{https://arxiv.org/abs/1104.5487}{{\tt 1104.5487}}].

\bibitem{Wainwright:2012zn}
C.~L. Wainwright, S.~Profumo, and M.~J. Ramsey-Musolf, {\it {Phase Transitions
  and Gauge Artifacts in an Abelian Higgs Plus Singlet Model}},  {\em Phys.
  Rev.} {\bf D86} (2012) 083537, [\href{https://arxiv.org/abs/1204.5464}{{\tt
  1204.5464}}].

\bibitem{Blinov:2015vma}
N.~Blinov, S.~Profumo, and T.~Stefaniak, {\it {The Electroweak Phase Transition
  in the Inert Doublet Model}},  {\em JCAP} {\bf 1507} (2015), no.~07 028,
  [\href{https://arxiv.org/abs/1504.05949}{{\tt 1504.05949}}].

\bibitem{Croon:2018new}
D.~Croon and G.~White, {\it {Exotic Gravitational Wave Signatures from
  Simultaneous Phase Transitions}},  {\em JHEP} {\bf 05} (2018) 210,
  [\href{https://arxiv.org/abs/1803.05438}{{\tt 1803.05438}}].

\bibitem{Ivanov:2017zjq}
I.~P. Ivanov, {\it {CP-symmetry of order 4 and its consequences}},  {\em J.
  Phys. Conf. Ser.} {\bf 873} (2017), no.~1 012036,
  [\href{https://arxiv.org/abs/1702.07542}{{\tt 1702.07542}}].

\bibitem{Caprini:2015zlo}
C.~Caprini {\em et~al.}, {\it {Science with the space-based interferometer
  eLISA. II: Gravitational waves from cosmological phase transitions}},  {\em
  JCAP} {\bf 1604} (2016), no.~04 001,
  [\href{https://arxiv.org/abs/1512.06239}{{\tt 1512.06239}}].

\bibitem{Caprini:2009yp}
C.~Caprini, R.~Durrer, and G.~Servant, {\it {The stochastic gravitational wave
  background from turbulence and magnetic fields generated by a first-order
  phase transition}},  {\em JCAP} {\bf 0912} (2009) 024,
  [\href{https://arxiv.org/abs/0909.0622}{{\tt 0909.0622}}].

\bibitem{Hindmarsh:2017gnf}
M.~Hindmarsh, S.~J. Huber, K.~Rummukainen, and D.~J. Weir, {\it {Shape of the
  acoustic gravitational wave power spectrum from a first order phase
  transition}},  {\em Phys. Rev.} {\bf D96} (2017), no.~10 103520,
  [\href{https://arxiv.org/abs/1704.05871}{{\tt 1704.05871}}].

\bibitem{Ellis:2019oqb}
J.~Ellis, M.~Lewicki, J.~M. No, and V.~Vaskonen, {\it {Gravitational wave
  energy budget in strongly supercooled phase transitions}},  {\em JCAP} {\bf
  1906} (2019), no.~06 024, [\href{https://arxiv.org/abs/1903.09642}{{\tt
  1903.09642}}].

\bibitem{Caprini:2019egz}
C.~Caprini {\em et~al.}, {\it {Detecting gravitational waves from cosmological
  phase transitions with LISA: an update}},
  \href{https://arxiv.org/abs/1910.13125}{{\tt 1910.13125}}.

\bibitem{Leitao:2015fmj}
L.~Leitao and A.~Megevand, {\it {Gravitational waves from a very strong
  electroweak phase transition}},  {\em JCAP} {\bf 1605} (2016), no.~05 037,
  [\href{https://arxiv.org/abs/1512.08962}{{\tt 1512.08962}}].

\bibitem{Espinosa:2010hh}
J.~R. Espinosa, T.~Konstandin, J.~M. No, and G.~Servant, {\it {Energy Budget of
  Cosmological First-order Phase Transitions}},  {\em JCAP} {\bf 1006} (2010)
  028, [\href{https://arxiv.org/abs/1004.4187}{{\tt 1004.4187}}].

\bibitem{Buonanno:2004tp}
A.~Buonanno, G.~Sigl, G.~G. Raffelt, H.-T. Janka, and E.~Muller, {\it
  {Stochastic gravitational wave background from cosmological supernovae}},
  {\em Phys. Rev.} {\bf D72} (2005) 084001,
  [\href{https://arxiv.org/abs/astro-ph/0412277}{{\tt astro-ph/0412277}}].

\bibitem{Thrane:2013oya}
E.~Thrane and J.~D. Romano, {\it {Sensitivity curves for searches for
  gravitational-wave backgrounds}},  {\em Phys. Rev.} {\bf D88} (2013), no.~12
  124032, [\href{https://arxiv.org/abs/1310.5300}{{\tt 1310.5300}}].

\bibitem{Moore:2014lga}
C.~J. Moore, R.~H. Cole, and C.~P.~L. Berry, {\it {Gravitational-wave
  sensitivity curves}},  {\em Class. Quant. Grav.} {\bf 32} (2015), no.~1
  015014, [\href{https://arxiv.org/abs/1408.0740}{{\tt 1408.0740}}].

\bibitem{Nakayama:2009ce}
K.~Nakayama and J.~Yokoyama, {\it {Gravitational Wave Background and
  Non-Gaussianity as a Probe of the Curvaton Scenario}},  {\em JCAP} {\bf 1001}
  (2010) 010, [\href{https://arxiv.org/abs/0910.0715}{{\tt 0910.0715}}].

\bibitem{Wan:2018udw}
Y.~Wan, B.~Imtiaz, and Y.-F. Cai, {\it {Cosmological phase transitions and
  gravitational waves in the singlet Majoron model}},
  \href{https://arxiv.org/abs/1804.05835}{{\tt 1804.05835}}.

\end{thebibliography}\endgroup

\end{document}